\documentclass{article}%
\usepackage{amsfonts}
\usepackage{amsmath}
\usepackage{amssymb}
\usepackage{graphicx}%
\providecommand{\U}[1]{\protect\rule{.1in}{.1in}}

\begin{document}

\title{Quantum Mechanics Associated with a Finite Group}
\author{Robert W. Johnson\\rwjcontact@aol.com}
\date{April 10, 2006}
\maketitle

\begin{abstract}
I describe, in the simplified context of finite groups and their
representations, a mathematical model for a physical system that contains both
its quantum and classical aspects. \ The physically observable system is
associated with the space containing elements $f\otimes f$ for $f$ an element
in the regular representation of a given finite group $G$. \ The Hermitian
portion of $f\otimes f$ is the Wigner distribution of $f$ whose convolution
with a test function leads to a mathematical description of the quantum
measurement process. \ Starting with the Jacobi group that is formed from the
semidirect product of the Heisenberg group with its automorphism group
$SL(2,\mathbb{F}_{N})$ for $N$ an odd prime number I show that the classical
phase space is the first order term in a series of subspaces of the Hermitian
portion of $f\otimes f$ that are stable under $SL(2,\mathbb{F}_{N}).$ \ I
define a derivative that is analogous to a pseudodifferential operator to
enable a treatment that parallels the continuum case. \ I give a new
derivation of the Schr\"{o}dinger-Weil representation of the Jacobi group.
\ Keywords: \ quantum mechamics, finite group, metaplectic. \ PACS:
\ 03.65.Fd; 02.10.De; 03.65.Ta.

\end{abstract}

\section{Introduction}

We consider the following construction: \ For a given finite group $G$:

\begin{enumerate}
\item Form the group ring over the complex numbers.

\item Decompose the group ring into subspaces $V_{i}$ that are invariant under
$G.$

\item For a vector\ $f$ in subspace $V_{i}$ form the tensor product $f\otimes
f.$

\item Decompose $f\otimes f$ into components that are invariant under the
action of $\{g\otimes g\mid$ $\ g\in S\subset G\}.$
\end{enumerate}

The main idea that we explore is the hypothesis that physical observables
reside in the vector space $f\otimes f.$ \ This vector space is closed under
operations $g\otimes g$, $g\in G$ that take $f\otimes f$ to $gf\otimes gf$ .
\ These operations compose, by hypothesis, the allowed transformations of the
corresponding physical observables.

In particular, we consider this construction for the semidirect product of the
Heisenberg group $\ H_{1}(\mathbb{F}_{N})$ with the special linear group
$SL(2,\mathbb{F}_{N}).$ $\ H_{1}(\mathbb{F}_{N})$ is generated by 2 elements
whose commutator is an element of the center of the group. \ The special
linear group $SL(2,\mathbb{F}_{N})$ is the group of automorphisms of
$H_{1}(\mathbb{F}_{N})$ that leaves the center invariant. $\ SL(2,\mathbb{F}%
_{N})$ is isomorphic to the symplectic group $SP(1,\mathbb{F}_{N}).$ \ \ The
semi-direct product $G^{J}=SL(2,\mathbb{F}_{N})\varpropto H_{1}(\mathbb{F}%
_{N})$ is the Jacobi group (Berndt and Schmidt 1998). \ We work over the prime
field $\mathbb{F}_{N}$ for $N$ a prime not equal to $2$. \ 

By working out the above construction for the Jacobi group $G^{J}$ we develop
a finite model for the quantum system comprised of a single particle with a
single spatial degree of freedom whose phase space is a flat torus. \ The
correspondence between the quantum system and our construction is the
following: \ The wavefunction for the quantum system corresponds to a vector
$f$ in an invariant subspace of $G^{J}$. \ The Hermitian component of
$f\otimes f$ corresponds to the Wigner distribution of $f$. \ The evaluation
at the origin of the convolution of the Hermitian component of\ $f\otimes f$
with a test function corresponds to the expectation value of the test
function. \ We obtain the Weyl map that relates the evaluation of a function
over the Wigner distribution with the expectation value of an operator. \ The
transformation of $f\otimes f$ under a 1 parameter family of operations
$g\otimes g$ for $\ g\in SL(2,\mathbb{F}_{N})\subset G^{J}$ corresponds to the
evolution of the quantum system under a homogeneous quadratic Hamiltonian.
\ We find subspaces of the Hermitian portion of $f\otimes f$ that are
invariant under $g\otimes g$ for $\ g\in SL(2,\mathbb{F}_{N})\subset G^{J}$.
\ These subspaces correspond to orders in the expansion of an exponential
operator expression. \ The zeroth order subspace corresponds to the quantum
total probability; the first order subspace corresponds to the classical phase
space. \ Higher order subspaces are also constructed. \ The quadratic forms
for the 2nd and 4th order subspaces correspond to lesser known "universal
quantum invariants" described by Dodonov 2000 and Dodonov and Man'ko 2000.

The Jacobi group over the real numbers is known in other works as the
Schr\"{o}dinger or Weyl-symplectic group (see, e.g., Guerrero et. al. 1999 and
references cited therein and Miller 1977) and also as the inhomogeneous
metaplectic group (de Gossen 2001). \ The problem of quantizing the classical
system whose phase space is a torus has been widely studied. \ Guerrero et.
al. apply the Algebraic Quantization on a Group approach (Aldaya 1989) to the
Jacobi (Schr\"{o}dinger) Group. \ Rieffel 1989 describes the deformation
quantization of toric symplectic manifolds. \ 

Quantum mechanics over a finite field\ began with work of Weyl and was
continued by Schwinger from the viewpoint of approximating quantum systems in
infinite dimensional spaces by those associated with finite abelian groups.
\ Varadarajan 1995 reviews this work and links it to the area of deformation
quantization. \ 

Athanasiu and Floratos 1994 consider the problem of the classical and quantum
evolution on phase space lattices having the structure of $N\otimes N$ for $N$
an odd prime. \ They start with the classical phase space with 1 degree of
freedom and consider the set of all linear canonical transformations acting on
this space. \ These form the group $SL(2,\mathbb{F}_{N})$. \ From the
classical phase space they pass to the Heisenberg group and its
Schr\"{o}dinger representation to construct the metaplectic representation of
$SL(2,\mathbb{F}_{N}).$ \ The metaplectic representation of $SL(2,\mathbb{F}%
_{N})$ is then used to describe the evolution of the quantum wavefunction.
\ The present paper incorporates the main ingredients of Athanasiu and
Floratos in a more general and rigorous setting. \ Rather than taking the
classical phase space as the starting point we take as starting point the
Jacobi group. \ The classical evolution that these authors start with
corresponds to the evolution of the first order subspace in a series of
subspaces that we derive.

In order to reach the goals of this paper, we use the discrete Fourier
transform (section \ref{disc FT}) to define a derivative operation (section
\ref{DerivDef}) acting on a function contained in the group algebra of a
cyclic group. \ This allows a treatment in the present discrete case that
parallels the continuum case. \ Our definition corresponds to the definition
of a pseudodifferential operator (Folland 1989 p93). \ 

We construct the Schr\"{o}dinger-Weil representation of the Jacobi group
(section \ref{JacobiRep}) that has the property that the action of operators
$g\in H_{1}(\mathbb{F}_{N})\subset G^{J}$coincides with the Schr\"{o}dinger
representation of $H_{1}(\mathbb{F}_{N})$ (see, e.g., Terras 1999 and
Grassberger and H\"{o}rmann 2001) and the action of $g\in SL(2,\mathbb{F}%
_{N})\subset G^{J}$ coincides with the metaplectic representation of
$SL(2,\mathbb{F}_{N})$ (Neuhauser 2002). \ Additional irreducible
representations of the Jacobi group can be formed from the tensor products of
the Schr\"{o}dinger-Weil representation with the representations for which the
Heisenberg group acts trivially and $SL(2,\mathbb{F}_{N})$ acts according to
one of its representations (Berndt and Schmidt 1998). \ These tensor product
representations, though also of interest, are not needed for our chosen model
system and will not be considered in this paper.

In section \ref{direct product} we consider the direct product $f\otimes f$
\ and show in section \ref{section fxf} that $P_{+}f\otimes f$ corresponds to
the Wigner distribution of $f.$ \ We next consider the convolution of
$P_{+}f\otimes f$ \ with a test function $\sigma_{G}$. \ For $f$ appropriately
normalized the value of the convolution $\sigma_{G}P_{+}f\otimes f$ at the
origin\ is the average value of the test function for the system\ $P_{+}%
f\otimes f.$ \ In section \ref{evaluation of test functions for P+fxf} we show
that this leads to the Weyl map that relates the expectation value of an
operator with the expectation value of a function evaluated over the Wigner distribution.

We consider the transformation of $P_{+}f\otimes f$ under $g\otimes g$ for
$g\in SL(2,\mathbb{F}_{N})$ in section \ref{transformation under sl2} and find
subspaces of $P_{+}f\otimes f$ that are stable under these transformations.
\ These subspaces correspond to orders in a power series expansion of
$P_{+}f\otimes f.$ \ We provide explicit representations of $SL(2,\mathbb{F}%
_{N})$ acting on these subspaces in sections \ref{zero order} to
\ref{fourth order} and give their associated invariant bilinear forms.

In order to make explicit contact with two familiar quantum systems in section
\ref{examples} we briefly treat the case of a free particle and the case of
the simple harmonic oscillator.

\ In this paper we apply the construction outlined in the first paragraph only
to the Jacobi group. \ This construction can in principle be applied to an
arbitrary finite group. \ This work extends similar constructions for
particular finite groups in Johnson 1996.

\bigskip

\section{\label{jacobi group}The Finite Jacobi Group}

\bigskip

The Jacobi group $G^{J}(\mathbb{F}_{N})$ for prime field $\mathbb{F}_{N}$ for
$N$ a prime number not equal to 2, is formed by the semi-direct product of the
Heisenberg Group $H_{1}(\mathbb{F}_{N})$\ and the special linear group
$SL(2,\mathbb{F}_{N})$: $\ G^{J}=SL(2,\mathbb{F}_{N})\varpropto H_{1}%
(\mathbb{F}_{N}).$ \ A convenient matrix representation is given by Berndt and
Schmidt 1998%
\begin{equation}
\left(
\begin{array}
[c]{cccc}%
a & 0 & b & \mu^{\prime}\\
\lambda & 1 & \mu & \kappa\\
c & 0 & d & -\lambda^{\prime}\\
0 & 0 & 0 & 1
\end{array}
\right)  \text{ with }ad-bc=1\text{ and }(\lambda,\mu)=(\lambda^{\prime}%
,\mu^{\prime})\left(
\begin{array}
[c]{cc}%
a & b\\
c & d
\end{array}
\right)  \qquad
\end{equation}
for a general element of $G^{J}$.

We identify
\begin{equation}
\left(
\begin{array}
[c]{cc}%
a & b\\
c & d
\end{array}
\right)  \in SL(2,\mathbb{F}_{N})\text{ with }\left(
\begin{array}
[c]{cccc}%
a & 0 & b & 0\\
0 & 1 & 0 & 0\\
c & 0 & d & 0\\
0 & 0 & 0 & 1
\end{array}
\right)  \in G^{J}%
\end{equation}
and
\begin{equation}
(\lambda,\mu,\kappa)\in H_{1}(\mathbb{F}_{N})\text{ with }\left(
\begin{array}
[c]{cccc}%
1 & 0 & 0 & \mu\\
\lambda & 1 & \mu & \kappa\\
0 & 0 & 1 & -\lambda\\
0 & 0 & 0 & 1
\end{array}
\right)  \in G^{J}%
\end{equation}
The Heisenberg group has multiplication
\begin{equation}
(\lambda,\mu,\kappa)(\lambda^{\prime},\mu^{\prime},\kappa^{\prime}%
)=(\lambda+\lambda^{\prime},\mu+\mu^{\prime},\kappa+\kappa^{\prime}+\lambda
\mu^{\prime}-\mu\lambda^{\prime})
\end{equation}
This parametrization can be translated to that of Folland (Folland 1989 p19)
by the change of variables $(p=\lambda,q=\mu,t=\kappa/2).$ with multiplication
$(p,q,t)(p^{\prime},q^{\prime},t^{\prime})=(p+p^{\prime},q+q^{\prime
},t+t^{\prime}+\frac{1}{2}(pq^{\prime}-qp^{\prime}))$. \ 

It will be useful in the following to consider a parametrization of the
Heisenberg group in terms of generators $t_{x}^{r},t_{y}^{s},t_{z}^{t}$
\begin{equation}
t_{x}^{r}=\left(
\begin{array}
[c]{cccc}%
1 & 0 & 0 & 0\\
r & 1 & 0 & 0\\
0 & 0 & 1 & -r\\
0 & 0 & 0 & 1
\end{array}
\right)  ;t_{y}^{s}=\left(
\begin{array}
[c]{cccc}%
1 & 0 & 0 & s\\
0 & 1 & s & 0\\
0 & 0 & 1 & 0\\
0 & 0 & 0 & 1
\end{array}
\right)  ;t_{z}^{t}=\left(
\begin{array}
[c]{cccc}%
1 & 0 & 0 & 0\\
0 & 1 & 0 & t\\
0 & 0 & 1 & 0\\
0 & 0 & 0 & 1
\end{array}
\right)
\end{equation}
for $r,s,t\in\mathbb{F}_{N}.$ \ A general element of the Heisenberg group can
then be written%

\begin{equation}
t_{x}^{r}t_{y}^{s}t_{z}^{t}=(r,s,t+rs).
\end{equation}
The group multiplication takes the form
\begin{equation}
\left(  t_{x}^{r}t_{y}^{s}t_{z}^{t}\right)  \left(  t_{x}^{r^{\prime}}%
t_{y}^{s^{\prime}}t_{z}^{t^{\prime}}\right)  =t_{x}^{r+r^{\prime}}%
t_{y}^{s+s^{\prime}}t_{z}^{t+t^{\prime}-2sr^{\prime}}%
\end{equation}
and we have the commutation relation
\begin{equation}
t_{x}^{r}t_{y}^{s}=t_{y}^{s}t_{x}^{r}t_{z}^{2rs}%
\end{equation}

The following (overcomplete) set of generators for $SL(2,\mathbb{F}%
_{N})=\{\left(
\begin{array}
[c]{cc}%
a & b\\
c & d
\end{array}
\right)  :ad-bd=1\}$ will be useful below:
\begin{align}
J  &  =\left(
\begin{array}
[c]{cc}%
0 & 1\\
-1 & 0
\end{array}
\right)  ,\text{ \ \ }t_{s}^{a}=\left(
\begin{array}
[c]{cc}%
a & 0\\
0 & \frac{1}{a}%
\end{array}
\right)  \text{ for }a\neq0\nonumber\\
\text{\ \ }t_{u}^{b}  &  =Jt_{d}^{-b}J^{-1}=\left(
\begin{array}
[c]{cc}%
1 & b\\
0 & 1
\end{array}
\right)  ,\text{ }t_{d}^{c}=\left(
\begin{array}
[c]{cc}%
1 & 0\\
c & 1
\end{array}
\right)  \text{ for }b,c\in\mathbb{F}_{N} \label{sl2std}%
\end{align}
\qquad\qquad with multiplication
\begin{align}
\ t_{s}^{a_{1}}t_{s}^{a_{2}}  &  =t_{s}^{a_{1}a_{2}},\text{ }t_{d}^{c_{1}%
}t_{d}^{c_{2}}=t_{d}^{c_{1}+c_{2}},\text{ }J^{4}=1\\
\ t_{s}^{a}t_{d}^{c}  &  =t_{d}^{c/a^{2}}t_{s}^{a},\text{ }\ t_{s}^{a}%
t_{u}^{b}=t_{u}^{ba^{2}}t_{s}^{a},Jt_{s}^{a}=t_{s}^{1/a}J,\text{ and,}\\
Jt_{d}^{c}  &  =t_{s}^{c}t_{d}^{-c}J^{-1}t_{d}^{-\frac{1}{c}}J\text{ for
}c\neq0\\
Jt_{u}^{b}  &  =t_{s}^{1/b}t_{u}^{-b}J^{-1}t_{u}^{-1/b}J^{-1}\text{ for }%
b\neq0 \label{Jtu}%
\end{align}
We note that $J^{2}=t_{s}^{-1}.$ \ The automorphisms of the Heisenberg group
that are induced by the $SL(2,\mathbb{F}_{N})$ generators are$:$
\begin{align}
J\left(  t_{x}^{r}t_{y}^{s}t_{z}^{t}\right)  J^{-1}  &  =t_{x}^{s}t_{y}%
^{-r}t_{z}^{t+2rs}\\
t_{s}^{a}\left(  t_{x}^{r}t_{y}^{s}t_{z}^{t}\right)  t_{s}^{1/a}  &
=t_{x}^{r/a}t_{y}^{as}t_{z}^{t}\\
t_{d}^{c}\left(  t_{x}^{r}t_{y}^{s}t_{z}^{t}\right)  t_{d}^{-c}  &
=t_{x}^{r-cs}t_{y}^{s}t_{z}^{t+cs^{2}}\\
t_{u}^{b}\left(  t_{x}^{r}t_{y}^{s}t_{z}^{t}\right)  t_{u}^{-b}  &  =t_{x}%
^{r}t_{y}^{s-br}t_{z}^{t+br^{2}}%
\end{align}

\subsection{Preliminaries}

\subsubsection{\label{rep theory}Representation theory for cyclic group of
order N}

In treating the Jacobi group we will deal with several abelian subgroups that
are generated by powers of a single generator. \ These include the subgroups
$\{t_{x}^{k}\},$ $\{t_{y}^{k}\},$ $\{t_{z}^{k}\},$ $\{t_{u}^{k}\},$ and
$\{Jt_{u}^{k}J^{-1}\}$ for $k\in\mathbb{F}_{N}.$ \ For each of these subgroups
we form invariant subspaces and irreducible representations in the same way.
\ In order to establish notation, let us review the representation theory for
a cyclic group $G$ of order $N$ (see, e.g., Terras 1999)%

\[
G=\{g^{k}\mid g^{N}=1,k\in\mathbb{F}_{N}\}
\]
We form the group algebra over the complex numbers with representative
element
\[
f=\sum_{k=0}^{N-1}f(k)g^{k}%
\]
for $f(k)$ a complex number. \ The element
\begin{equation}
\widehat{g}_{m}=\sum_{k=0}^{N-1}\exp(\frac{-2\pi i}{N}mk)g^{k},
\end{equation}
determines a $1$ dimensional invariant subspace in the group algebra for each
$m\in\mathbb{F}_{N}$ . \ We have
\begin{equation}
g^{a}\widehat{g}_{m}=\exp(\frac{2\pi i}{N}ma)\widehat{g}_{m}%
\end{equation}
for $g^{a}\in G.$ \ The eigenvalue $\exp(\frac{2\pi i}{N}ma)$ is the character
of $g^{a}$ in the representation determined by the action of $G$ on
$\widehat{g}_{m}.$ \ We also have
\begin{align}
\widehat{g}_{m}\widehat{g}_{m^{\prime}}  &  =N\widehat{g}_{m}\delta
(m-m^{\prime})\\
\frac{1}{N}\sum_{m=0}^{N-1}\widehat{g}_{m}  &  =1
\end{align}

\subsubsection{\label{disc FT}Discrete Fourier Transform}

Acting on a general element $f$ in the group algebra
\begin{equation}
f=\sum_{k=0}^{N-1}f(k)g^{k} \label{f}%
\end{equation}
with $\frac{1}{N}\sum_{m=0}^{N-1}\widehat{g}_{m}=1$ we obtain
\begin{align}
f  &  =\frac{1}{N}\sum_{m,k=0}^{N-1}f(k)\exp(\frac{2\pi i}{N}mk)\widehat
{g}_{m}\label{gsubm}\\
&  =\sum_{m=0}^{N-1}\widetilde{f}(m)\widehat{g}_{m}%
\end{align}
where
\begin{equation}
\widetilde{f}(m)=\frac{1}{N}\sum_{k=0}^{N-1}f(k)\exp(\frac{2\pi i}{N}mk)
\end{equation}
is the inverse discrete Fourier transform of $f(k)$(see, e.g., Bachman et. al.
2000). \ Now expanding out $\widehat{g}_{m}$ in equation \ref{gsubm}
\begin{equation}
f=\sum_{k,m=0}^{N-1}\widetilde{f}(m)\exp(\frac{-2\pi i}{p}mk)g^{k}%
\end{equation}
and comparing this series with equation \ref{f} we obtain the discrete Fourier
transform
\begin{equation}
f(k)=\sum_{m=0}^{N-1}\widetilde{f}(m)\exp(\frac{-2\pi i}{N}mk)
\end{equation}
\qquad\qquad\ 

\subsubsection{\label{DerivDef}Derivative operation for finite cyclic group}

We use the discrete Fourier transform of $f(k)$ to define a derivative
operation and power series expansion of $f(k)$. \ \ In analogy with the case
for $k$ a real number we define
\begin{align}
&  \frac{d^{n}}{dk^{n}}f(k)\overset{def}{=}\sum_{m=0}^{N-1}(\frac{-2\pi i}%
{N}m)^{n}\widetilde{f}(m)\exp(\frac{-2\pi i}{N}mk)\\
&  =\frac{1}{N}\sum_{h,m=0}^{N-1}(\frac{-2\pi i}{N}m)^{n}f(h)\exp(\frac{2\pi
i}{N}m(h-k)) \label{deriv}%
\end{align}
This definition parallels that used to define a pseudodifferential operator
(Folland 1989 p93).

Consider the power series expansion of the exponential
\begin{equation}
\exp\left(  l\frac{d}{dk}\right)  f(k)=\sum_{n=0}^{\infty}\frac{l^{n}}%
{n!}\left(  \frac{d^{n}}{dk^{n}}\right)  f(k)
\end{equation}
Using equation \ref{deriv} for the $n^{th}$ order derivative we can express
$f(k+l)$ in terms of the exponential of a derivative operation:%

\begin{align}
\exp\left(  l\frac{d}{dk}\right)  f(k)  &  =\sum_{h,m=0}^{N-1}\sum
_{n=0}^{\infty}\frac{l^{n}}{n!}\frac{1}{N}(\frac{-2\pi i}{N}m)^{n}%
f(h)\exp(\frac{2\pi i}{N}m(h-k)))\nonumber\\
&  =\sum_{h,m=0}^{N-1}\frac{1}{N}f(h)\exp(\frac{2\pi i}{N}m(h-k-l))\nonumber\\
&  =f(k+l). \label{f(k+h)}%
\end{align}
Applying this formula to the function $f(k)=k^{n}$ and identifying terms
having the same order of $l$ on both sides of the equation we obtain, e.g.,
$\frac{d}{dk}k^{n}=nk^{n-1}.$ \ Applying this formula to a function formed
from the product of 2 functions
\begin{align*}
\exp\left(  l\frac{d}{dk}\right)  f_{1}(k)f_{2}(k)  &  =f_{1}(k+l)f_{2}(k+l)\\
&  =\left(  \exp\left(  l\frac{d}{dk}\right)  f_{1}(k)\right)  \left(
\exp\left(  l\frac{d}{dk}\right)  f_{2}(k)\right)
\end{align*}
expanding out the exponentials on both sides of the equality, and identifying
the first order terms in $l$ we obtain the Leibnitz property. \ We then also
obtain the commutator expression
\[
\lbrack\frac{d}{dk},k]=1.
\]
Using $\sum_{k=0}^{N-1}\frac{d}{dk}\left[  x(k)y(k)\right]  =0\ $that follows
from equation \ref{deriv} we also have,
\begin{equation}
\sum_{k=0}^{N-1}\left(  \frac{d}{dk}x(k)\right)  y(k)=-\sum_{k=0}%
^{N-1}x(k)\frac{d}{dk}y(k)
\end{equation}
and
\begin{equation}
\sum_{k=0}^{N-1}\left(  \frac{d^{n}}{dk^{n}}x(k)\right)  y(k)=\left(
-1\right)  ^{n}\sum_{k=0}^{N-1}x(k)\frac{d^{n}}{dk^{n}}y(k) \label{deriv rev}%
\end{equation}

\subsection{\bigskip\label{HeisRep}Representations of the Heisenberg Group}

The representation theory for the Heisenberg group over a finite commutative
ring is presented in Terras 1999 and Grassberger and H\"{o}rmann 2001. \ We
consider the regular representation of $H_{1}(\mathbb{F}_{N})$ over the
complex numbers with general element%

\[
f=\sum_{r,s,t=0}^{N-1}f(r,s,t)t_{x}^{r}t_{y}^{s}t_{z}^{t}%
\]
for $f(r,s,t)$ a complex number. \ The subgroup generated by $\{t_{y}%
^{s},t_{z}^{t}\mid s,t\in\mathbb{F}_{N}\}$ is abelian. \ $H_{1}(\mathbb{F}%
_{N})$ is the semidirect product of this subgroup with the subgroup generated
by $\{t_{x}^{r}\mid r\in\mathbb{F}_{N}\}$. \ We first form invariant subspaces
in the $\{t_{y}^{s},t_{z}^{t}\}$ group algebra. \ The product $\widehat
{y}_{\nu}\widehat{z}_{\omega}$ for
\begin{align*}
\widehat{y}_{\nu}  &  =\sum_{n=0}^{N-1}\exp(\frac{-2\pi i}{N}\nu n)t_{y}^{n}\\
\widehat{z}_{\omega}  &  =\ \sum_{m=0}^{N-1}\exp(\frac{-2\pi i}{N}\omega
m)t_{z}^{m}%
\end{align*}
and $\nu,\omega\in\mathbb{F}_{N}$ is invariant under the $\{t_{y}^{r}%
,t_{z}^{t}\}$ subgroup$.$ \ Acting on the left of $\ \widehat{y}_{\nu}%
\widehat{z}_{\omega}$ with a general element of the Heisenberg group algebra
we obtain
\begin{equation}
f_{\nu\omega}=\sum_{k=0}^{N-1}f(k)t_{x}^{k}\widehat{y}_{\nu}\widehat
{z}_{\omega}%
\end{equation}
where $f(k)\in\mathbb{C}$. \ $f_{\nu\omega}$ is a general element in the left
ideal determined by $\widehat{y}_{\nu}\widehat{z}_{\omega}.$ \ The left action
of a general element $t_{x}^{r}t_{y}^{s}t_{z}^{t}\in H_{1}(\mathbb{F}_{N})$ on
$f_{\nu\omega}$ is given by
\begin{equation}
t_{x}^{r}t_{y}^{s}t_{z}^{t}\ f_{\nu\omega}=\sum_{k=0}^{N-1}\exp(\frac{2\pi
i}{N}(\omega t+s\nu-2\omega ks+2\omega rs))f(k-r)t_{x}^{k}\widehat{y}_{\nu
}\widehat{z}_{\omega}.\ \label{schrep}%
\end{equation}
For $\omega\neq0$ this representation corresponds to the Schr\"{o}dinger
representation of $t_{x}^{r}t_{y}^{s}t_{z}^{t}$ $\in H_{1}(\mathbb{F}_{N})$.
\ That this representation is irreducible and unique up to equivalence for
each $\omega\neq0$ is described in Grassberger and H\"{o}rmann 2001 and Terras 1999.

\subsection{\bigskip\label{JacobiRep}A Representation of the Jacobi Group}

We wish now to obtain a representation of the Jacobi Group $\ G^{J}%
=SL(2,\mathbb{F}_{N})\varpropto H_{1}(\mathbb{F}_{N})$ that behaves under the
Heisenberg subgroup $H_{1}(\mathbb{F}_{N})$ according to the Schr\"{o}dinger
representation. \ For this purpose, consider an element $f$ in the Jacobi
group algebra with the following form:
\begin{equation}
f=\sum_{k=0}^{N-1}f(k)t_{x}^{k}\widehat{y}_{0}\widehat{z}_{\omega}\widehat
{s}_{-}\widehat{u}_{0}\widehat{x}_{0}\left[  1+\frac{1}{\left(  \frac{\omega
}{N}\right)  G(1,N)}J\right]  \widehat{d}_{0} \label{jacrp}%
\end{equation}
\ In this expression $f(k)$ is a complex number,
\begin{equation}
\widehat{y}_{0}=\sum_{m=0}^{N-1}t_{y}^{m}%
\end{equation}
is the eigenstate of $t_{y}$with eigenvalue 1,
\begin{equation}
\widehat{z}_{\omega}=\sum_{h=0}^{N-1}exp(\frac{-2\pi i}{N}h\omega)t_{z}^{h}%
\end{equation}
is the eigenstate of $t_{z}^{t}$ with eigenvalue $exp(\frac{2\pi i}{N}%
t\omega),$
\begin{equation}
\widehat{s}_{-}=\sum_{m=1}^{N-1}\left(  \frac{m}{N}\right)  t_{s}^{m}%
\end{equation}
is the eigenstate of \ $t_{s}^{a}\ $with eigenvalue $\left(  \frac{a}%
{N}\right)  .$ \ In this expression $\left(  \frac{m}{N}\right)  $ is the
Legendre symbol. \ The Legendre symbol has value 1 for $m$ a square modulo $N$
and $-1$ for $m$ a nonsquare modulo $N$.
\begin{align}
\widehat{u}_{0}  &  =\sum_{k=0}^{N-1}t_{u}^{k}\\
\widehat{x}_{0}  &  =\sum_{h=0}^{N-1}t_{x}^{h},\\
\widehat{d}_{0}  &  =\sum_{l=0}^{N-1}t_{d}^{l},
\end{align}
and $\left(  \frac{\omega}{N}\right)  G(1,N)$ is the product of the Legendre
symbol $\left(  \frac{\omega}{N}\right)  $ with the Gauss sum (Lang 1994)
\begin{equation}
G(1,N)={\Huge \{}_{i\sqrt{N}for\text{ }N=3\operatorname{mod}4}^{\sqrt
{N}for\text{ }N=1\operatorname{mod}4}%
\end{equation}
\ Let us denote by $I$ the operator\textit{\ }
\begin{equation}
I=\widehat{y}_{0}\widehat{z}_{\omega}\widehat{s}_{-}\widehat{u}_{0}\widehat
{x}_{0}\left(  1+\frac{1}{\left(  \frac{\omega}{N}\right)  G(1,N)}J\right)
\widehat{d}_{0} \label{I eq}%
\end{equation}

Let us now verify that such an $f$ is invariant under $G^{J}$. \ By direct
calculation we find for the Heisenberg group elements $\{t_{x}^{r}t_{y}%
^{s}t_{z}^{t}\mid r,s,t\in\mathbb{F}_{N}\}$
\begin{equation}
t_{x}^{r}t_{y}^{s}t_{z}^{t}f=\sum_{k}\exp(\frac{2\pi i}{N}(\omega t-2\omega
ks+2\omega rs))f(k-r)t_{x}^{k}I
\end{equation}
This corresponds to the Schr\"{o}dinger representation equation \ref{schrep}
with $\upsilon=0$. \ For the $SL(2,\mathbb{F}_{N})$ group generators we find
\begin{align}
t_{s}^{a}f  &  =\left(  \frac{a}{N}\right)  \sum_{k=0}^{N-1}f(ak)t_{x}%
^{k}I\label{tsxts}\\
t_{u}^{b}f  &  =\sum_{k=0}^{N-1}\exp(\frac{2\pi i}{N}bk^{2}\omega
)f(k)t_{x}^{k}I\label{tuxtu}\\
Jf  &  =\frac{\left(  \frac{\omega}{N}\right)  G(1,N)}{N}\sum_{k,l=0}%
^{N-1}f(k)\exp(\frac{2\pi i}{N}2\omega kl)t_{x}^{l}I \label{JxJ}%
\end{align}
Equations \ref{tsxts} and \ref{tuxtu} are easily calculated since one needs
only to pass\ $t_{s}^{a}$ and $t_{u}^{b}$ through equation \ref{jacrp} until
$t_{s}^{a}$ is absorbed in $\widehat{s}_{-}$ and $t_{u}^{b}$ is absorbed in
$\widehat{u}_{0}$. \ The calculation for $Jf,$ though elementary, is involved
and I provide details in Appendix A. \ This representation for the
$SL(2,\mathbb{F}_{N})$ subgroup coincides with the metaplectic (or, Weil)
representation of $SL(2,\mathbb{F}_{N})$ (Neuhauser 2002). \ This is an
ordinary representation for $SL(2,\mathbb{F}_{N}).$ \ Neuhauser treats the
more general case of $SL(2,K)$ for $K$ a finite field; Cliff et. al.
2000\ \ consider the case for symplectic groups over rings. \ The derivation
of the metaplectic representation through the action of $SL(2,\mathbb{F}_{N})$
on an ideal in the Jacobi group algebra that I give above is new.

This representation for the Jacobi group is known as the Schr\"{o}dinger-Weil
representation. \ As described by Berndt and Schmidt 1998 for the real,
complex and p-adic cases, additional representations of the Jacobi group can
now be obtained by forming the tensor product of the Schr\"{o}dinger-Weil
representation with representations in which the Heisenberg subgroup acts
trivially and the $SL(2)$ subgroup acts according to one of its
representations. \ These representations, though also of interest, are not
needed for the description of the quantum system that I model here and will
not be considered in this paper.

It will be useful to have an expression available for the action of $t_{d}%
^{c}$ on $f$%
\begin{align}
t_{d}^{c}f  &  =Jt_{u}^{-c}t_{s}^{-1}Jf=\sum_{q=0}^{N-1}\exp(\frac{-2\pi i}%
{N}\frac{cq^{2}}{4\omega})\widetilde{f}(q)\widehat{x}_{q}I\label{tdf}\\
&  =\sum_{k}^{N-1}\exp(\frac{-N}{2\pi i}\frac{c}{4\omega}\frac{d^{2}}{dk^{2}%
})f(k)t_{x}^{k}I.\nonumber
\end{align}
One may verify the second expression by expanding out the exponential and
using the definition equation \ref{deriv} for the derivative.

The parity operator $J^{2}$ commutes with $SL(2,\mathbb{F}_{N})$, intertwines
with $t_{x}^{r}t_{y}^{s}\in H_{1}(\mathbb{F}_{N})$ changing the sign of the
translation $J^{2}t_{x}^{r}t_{y}^{s}=t_{x}^{-r}t_{y}^{-s}J^{2},$ and has
square equal to 1. \ We note but will not use the property that the operator
\begin{equation}
P=\frac{1}{N}\sum_{k=0}^{N-1}\widehat{y}_{-k\omega}t_{x}^{k}%
\end{equation}
acts just as $J^{2}$ within the above ideal of $G^{J}.$

\section{\label{direct product}\bigskip Direct product algebra}

\subsubsection{Cyclic group of order 4}

As a first simple example that will also be useful in the following, let us
apply the construction outlined in the introduction to the cyclic group of
order 4 $C_{4}=\{1,g,g^{2},g^{3}\mid g^{4}=1\}.$ \ The group ring over the
real numbers contains elements
\begin{equation}
a=a_{0}1+a_{1}g+a_{2}g^{2}+a_{3}g^{3}%
\end{equation}
for $a_{i}\in\operatorname{real}$ numbers. \ We decompose this algebra into
subspaces using the projection operators $p_{\pm}=\frac{1}{2}(1\pm g^{2})$%
\begin{align}
a  &  =\left[  (a_{0}+a_{2})1+(a_{1}+a_{3})g\right]  p_{+}\nonumber\\
&  +\left[  (a_{0}-a_{2})1+(a_{1}-a_{3})g\right]  p_{-}%
\end{align}
$g^{2}$ acts trivially on the $p_{+}$ subspace and it can be further
decomposed into 2 1-dimensional subspaces that are respectively symmetric and
antisymmetric under $g$. \ We will not consider these further$.$ \ The $p_{-}$
subspace that is antisymmetric for $g^{2}$ is 2-dimensional and irreducible
over the real numbers. \ Mapping $g$ to the unit imaginary $i$ we\ may
identify the $p_{-}$ subspace with the complex numbers.

Let us consider the direct product of 2 distinct elements $a,b$ in the $p_{-}$
subspace
\[
a\otimes b=(a_{0}+ia_{1})p_{-}\otimes(b_{0}+ib_{1})p_{-}%
\]
\ Let us introduce the notation $E=i\otimes i$ and form projection operators
$P_{\pm}=1/2(1\pm E).$ \ We may then write%
\begin{align*}
a\otimes b  &  ={\large \{}a_{0}\left(  1\otimes1\right)  -a_{1}E\left(
1\otimes i\right)  )(b_{0}\left(  1\otimes1\right)  +b_{1}\left(  1\otimes
i\right)  {\large \}}\left(  p_{-}\otimes p_{-}\right) \\
&  ={\LARGE \{}P_{+}\left[  (a_{0}b_{0}+a_{1}b_{1})\left(  1\otimes1\right)
+(a_{0}b_{1}-a_{1}b_{0})(1\otimes i)\right] \\
&  +P_{-}\left[  (a_{0}b_{0}-a_{1}b_{1})\left(  1\otimes1\right)  +(a_{0}%
b_{1}+a_{1}b_{0})(1\otimes i)\right]  {\LARGE \}}\left(  p_{-}\otimes
p_{-}\right)
\end{align*}
Let us identify $\left(  1\otimes i\right)  $ with the unit imaginary in the
direct product space. \ The $P_{-}$ subspace is invariant under the
interchange $a\otimes b\rightarrow b\otimes a$. \ This interchange corresponds
to complex conjugation of the $P_{+}$ subspace. \ We may identify the $P_{+}$
subspace with the Hermitian product of 2 complex numbers. \ The $P_{-}$
subspace corresponds to the ordinary nonhermitean product of 2 complex numbers.

The group algebra of $G^{J}$ over the complex numbers coincides with the
$p_{-}$ portion of the group algebra of $C_{4}\otimes G^{J}$over the real
numbers. \ For $x$ an element of this group algebra, $P_{+}$ $x\otimes x$
corresponds to the Hermitian portion of $x\otimes x.$ \ In the following we
will mainly be concerned with this Hermitian portion.

\subsubsection{\bigskip Useful operators in the direct product algebra}

We introduce the translation operators
\begin{align}
T_{x}^{k}  &  =t_{x}^{k}\otimes t_{x}^{k},\text{ \ }\overline{T}_{x}^{k}%
=t_{x}^{-k}\otimes t_{x}^{k}\\
T_{y}^{h}  &  =t_{y}^{h}\otimes t_{y}^{h},\text{ \ }\overline{T}_{y}^{h}%
=t_{y}^{-h}\otimes t_{y}^{h}%
\end{align}
and form eigenvectors
\begin{align}
X_{p}  &  =\underset{h=0}{\sum^{N-1}}\exp(\frac{-2\pi i}{N}ph)T_{x}^{h},\text{
\ \ }\overline{X}_{q}=\underset{h=0}{\sum^{N-1}}\exp(\frac{-2\pi i}%
{N}qh)\overline{T}_{x}^{h}\\
Y_{p}  &  =\underset{h=0}{\sum^{N-1}}\exp(\frac{-2\pi i}{N}ph)T_{y}^{h},\text{
\ \ }\overline{Y}_{q}=\underset{h=0}{\sum^{N-1}}\exp(\frac{-2\pi i}%
{N}qh)\overline{T}_{y}^{h}%
\end{align}
We calculate the commutators
\begin{align}
T_{x}^{r}T_{y}^{s}T_{x}^{-r}T_{y}^{-s}\left(  \widehat{z}_{\omega}%
\otimes\widehat{z}_{\omega}\right)   &  =\exp(\frac{2\pi i}{N}2rs\omega
(1-E))\left(  \widehat{z}_{\omega}\otimes\widehat{z}_{\omega}\right)
\label{comuTT}\\
T_{x}^{r}\overline{T}_{y}^{s}T_{x}^{-r}\overline{T}_{y}^{-s}\left(
\widehat{z}_{\omega}\otimes\widehat{z}_{\omega}\right)   &  =\exp(\frac{2\pi
i}{N}2rs\omega(1+E))\left(  \widehat{z}_{\omega}\otimes\widehat{z}_{\omega
}\right)  \label{comTTb}%
\end{align}

\subsubsection{Two useful identities}

\bigskip The result that we consider in this paragraph will help to
characterize $P_{+}\left(  f\otimes f\right)  $. \ We prove
\begin{gather}
\left(  \sum_{r,s=0}^{N-1}(-r)^{m}(-s)^{n}T_{x}^{r}T_{y}^{s}\right)  \left(
\sum_{q,p=0}^{N-1}q^{h}p^{k}X_{q}Y_{p}\right)  \mid_{0,0}=\label{test f id}\\
\text{ \ \ \ \ \ \ \ \ \ \ \ \ \ \ \ \ \ \ \ \ \ \ \ }=\left(  \frac{N}{2\pi
i}\right)  ^{m+n}N^{2}m!n!\delta(m-h)\delta(n-k)\nonumber
\end{gather}
where the expression on the left is evaluated at the origin of the group ring
of the group generated by $\{T_{x}^{r},T_{y}^{s}\mid r,s\in\mathbb{F}_{N}\}$.
\ Acting with $T_{x}^{r}T_{y}^{s}$ on the eigenvectors $X_{q}Y_{p}$ in the
left-hand side of equation \ref{test f id} pulls out an exponential term
$exp(\frac{2\pi i}{N}(rq+sp)).$ \ Expand out the eigenvectors $X_{q}Y_{p}$ and
consider the value of the expression at the origin $T_{x}^{0}T_{y}^{0}.$ \ We
obtain for the left hand side
\[
\sum_{r,s,q,p=0}^{N-1}(-r)^{m}(-s)^{n}q^{h}p^{k}\exp(\frac{2\pi i}{N}(rq+sp)
\]
We write this expression as
\[
=\sum_{r,s,q,p=0}^{N-1}q^{h}p^{k}\left(  \frac{-N}{2\pi i}\right)
^{m+n}\left(  \frac{d}{dq}\right)  ^{m}\left(  \frac{d}{dp}\right)  ^{n}%
\exp(\frac{2\pi i}{N}(rq+sp)
\]
and use equation \ref{deriv rev} to reexpress this in the form
\[
=\sum_{r,s,q,p=0}^{N-1}\left(  \frac{N}{2\pi i}\right)  ^{m+n}\exp(\frac{2\pi
i}{N}(rq+sp)\left(  \frac{d}{dq}\right)  ^{m}\left(  \frac{d}{dp}\right)
^{n}q^{h}p^{k}%
\]
For $m>h$ or $n>k$ the derivatives lead to a null result. \ For $m<h$ or $n<k
$ the sum over $r,s$ in the exponential leads to a null result. \ We obtain a
nonnull result only for $m=h$ and $n=k$. \ We conclude
\begin{multline}
\sum_{r,s,q,p=0}^{N-1}(-r)^{m}(-s)^{n}q^{h}p^{k}\exp(\frac{2\pi i}%
{N}(rq+sp)=\nonumber\\
=\left(  \frac{N}{2\pi i}\right)  ^{m+n}N^{2}m!n!\delta(m-h)\delta(n-k)
\end{multline}

In this paragraph we rewrite an operator expression that will allow us to
describe subspaces of $f\otimes f$ below in a more transparent way. \ We show
that
\begin{equation}
\overline{T}_{x}^{r}\left(  I\otimes I\right)  =(P_{+}Y_{-4\omega r}%
+P_{-}\overline{Y}_{-4\omega r})\overline{X}_{0}\left(  I\otimes I\right)  /N
\label{reexpTr/2}%
\end{equation}
for $I$ given by equation \ref{I eq}. Consider
\begin{align*}
P_{+}\overline{X}_{q}\left(  I\otimes I\right)   &  =P_{+}\overline{X}%
_{q}T_{y}^{k}\left(  I\otimes I\right) \\
&  =P_{+}T_{y}^{k}\overline{X}_{q-4k\omega}\left(  I\otimes I\right)
\end{align*}
For $q-4k\omega=0$%
\[
P_{+}\overline{X}_{q}\left(  I\otimes I\right)  =P_{+}T_{y}^{q/4\omega
}\overline{X}_{0}\left(  I\otimes I\right)  .
\]
We then have
\begin{align}
P_{+}\overline{T}_{x}^{r}\left(  I\otimes I\right)   &  =P_{+}\underset
{q=0}{\sum^{N-1}}\exp\left(  2\pi irq/N\right)  \overline{X}_{q}\left(
I\otimes I\right)  /N\nonumber\\
&  =P_{+}\underset{q=0}{\sum^{N-1}}\exp\left(  2\pi irq/N\right)
T_{y}^{q/4\omega}\overline{X}_{0}\left(  I\otimes I\right)  /N\nonumber\\
&  =P_{+}\text{ }Y_{-4\omega r}\overline{X}_{0}\left(  I\otimes I\right)  /N
\end{align}
Similarly, we find%
\[
P_{-}\overline{T}_{x}^{r}\left(  I\otimes I\right)  =P_{-}\overline
{Y}_{-4\omega r}\overline{X}_{0}\left(  I\otimes I\right)  /N
\]
Summarizing, we have equation \ref{reexpTr/2}.

\subsection{\label{section fxf}Direct product $f\otimes f$}

We take an element $f$ in the left ideal of the Jacobi group algebra,
\begin{equation}
f=\sum_{k=0}^{N-1}f(k)t_{x}^{k}I
\end{equation}
where $I=\widehat{y}_{0}\widehat{z}_{\omega}\widehat{s}_{-}\widehat{u}%
_{0}\widehat{x}_{0}\left(  1+\frac{1}{\left(  \frac{\omega}{N}\right)
G(1,N)}J\right)  \widehat{d}_{0}$ as described above, and form the direct
product of 2 copies of this element
\begin{align}
f\otimes f  &  =\left(  \sum_{k=0}^{N-1}f(k)t_{x}^{k}I\right)  \otimes\left(
\sum_{h=0}^{N-1}f(h)t_{x}^{h}I\right) \nonumber\\
&  =\underset{h,k=0}{\sum^{N-1}}\left(  f(k)\otimes f(h)\right)  \left(
t_{x}^{k}\otimes t_{x}^{h}\right)  \left(  I\otimes I\right) \nonumber\\
&  =\underset{k,p=0}{\sum^{N-1}}\left(  f(k)\otimes f(k+p)\right)  T_{x}%
^{k}\left(  1\otimes t_{x}^{p}\right)  \left(  I\otimes I\right)
\end{align}
where $p=h-k.$ \ Acting on $f\otimes f$ with $1=\underset{}{\frac{1}{N}%
\sum_{q}}X_{q}$ and using $T_{x}^{p/2}\overline{T}_{x}^{p/2}=1\otimes
t_{x}^{p}$ we obtain
\[
f\otimes f=\sum_{k,p,q=0}^{N-1}\left(  f(k)\otimes\exp(\frac{2\pi i}%
{N}q(k+p/2)f(k+p)\right)  \ X_{q}\overline{T}_{x}^{p/2}\left(  I\otimes
I\right)  /N
\]
We use equation \ref{f(k+h)} to express $f(k+p)$ in terms of the exponential
of a derivative operation
\[
f(k+p)=\exp\left(  p\frac{d}{dk}\right)  f(k)
\]
We then express $f\otimes f$ in the form
\begin{gather}
f\otimes f=\sum_{k,p,q=0}^{N-1}f(k)\otimes\exp(\frac{2\pi i}{N}q(k+p/2)\exp
\left(  p\frac{d}{dk}\right)  f(k)\text{ \ }\nonumber\\
\cdot X_{q}\overline{T}_{x}^{p/2}\left(  I\otimes I\right)  /N\nonumber\\
=\sum_{k,p,q=0}^{N-1}f(k)\otimes\exp\left(  \frac{2\pi i}{N}(qk+p\left(
\frac{N}{2\pi i}\frac{d}{dk}\right)  \right)  f(k)\text{ }\label{fxf}\\
\cdot\text{\ }X_{q}\overline{T}_{x}^{p/2}\left(  I\otimes I\right)
/N\nonumber
\end{gather}
where in the second equality we have used the Campbell-Hausdorf expression
\begin{equation}
\exp(A)\exp(B)=\exp(A+B)\exp(1/2[A,B])
\end{equation}
valid for the case that $A$ and $B\ $commute with their commutator. \ Finally,
substituting using equation \ref{reexpTr/2} we obtain
\begin{gather}
f\otimes f={\Huge \{}P_{+}\sum_{k,p,q=0}^{N-1}f(k)^{\ast}\exp\left(
\frac{2\pi i}{N}(qk+p\left(  \frac{-1}{2\omega}\frac{N}{2\pi i}\frac{d}%
{dk}\right)  \right)  f(k)\text{ \ }X_{q}Y_{p}+\nonumber\\
+P_{-}\sum_{k,p,q=0}^{N-1}f(k)\exp\left(  \frac{2\pi i}{N}(qk+p\left(
\frac{-1}{2\omega}\frac{N}{2\pi i}\frac{d}{dk}\right)  \right)  f(k)\text{
\ }X_{q}\overline{Y}_{p}{\Huge \}}\nonumber\\
\cdot\overline{X}_{0}\left(  I\otimes I\right)  /N^{2} \label{fwig}%
\end{gather}
We identify the $P_{+}$ portion of this expression with the Fourier-Wigner
distribution of $f$ (Folland 1989 p30). \ Note that in comparing with Folland,
our term $\left(  \frac{N}{2\pi i}\frac{d}{dk}\right)  $ corresponds to
Folland's $D$ and our term $\frac{-1}{2\omega}$ corresponds to Folland's
incorporation of the Heisenberg constant $h.$

We now rewrite equation \ref{fwig} in terms of the generators $\{T_{x}%
,T_{y}\}$ for the $P_{+}$ projection (this amounts to a double Fourier
transform) and in terms of the generators $\{T_{x},\overline{T}_{y}\} $ for
the $P_{-}$ projection
\begin{gather}
f\otimes f={\Huge \{}P_{+}\sum_{r,s,p=0}^{N-1}f(r-\left(  \frac{-1}{2\omega
}\right)  \frac{p}{2})^{\ast}f(r+\left(  \frac{-1}{2\omega}\right)  \frac
{p}{2})\exp(\frac{-2\pi i}{N}ps)\text{ }\ T_{x}^{r}T_{y}^{s}+\nonumber\\
+P_{-}\sum_{r,s,p=0}^{N-1}f(r-\left(  \frac{-1}{2\omega}\right)  \frac{p}%
{2})f(r+\left(  \frac{-1}{2\omega}\right)  \frac{p}{2})\exp(\frac{-2\pi i}%
{N}ps)\text{ }\ T_{x}^{t}\overline{T}_{y}^{s}{\Huge \}}\nonumber\\
\cdot\overline{X}_{0}\left(  I\otimes I\right)  /N \label{wigd}%
\end{gather}
We identify the $P_{+}$ $f\otimes f$ portion of this expression with the
Wigner distribution of $f$ (Folland 1989 p57). \ This identification is
central to the hypothesis outlined in the introduction that associates
physical observables with the space $f\otimes f$ .

\subsection{\bigskip\label{characterization of fxf}Characterization of
$f\otimes f$}

\subsubsection{\bigskip Convolution with Test Function}

Let us now focus on extracting information from $f\otimes f.$ \ Let us
specialize to the Hermitian $P_{+}$ $f\otimes f$ portion since it is of most
direct physical interest.

$P_{+}$ $f\otimes f$ in the Wigner form equation \ref{wigd} composes a
function on the regular representation of the abelian group (see, equation
\ref{comuTT})$\ \{T_{x}^{h}T_{y}^{s}\mid r,s\in\mathbb{F}_{N}\}$. \ \ To
motivate our viewpoint let us consider a specific test function
\begin{equation}
\sigma_{G}=\sum_{r,s=0}^{N-1}-rT_{x}^{r}T_{y}^{s}=\sum_{r=0}^{N-1}-rT_{x}%
^{r}Y_{0}%
\end{equation}
and consider the convolution of \ $\sigma_{G}$ with a function $F$ that is a
composed of contributions $a(h)$%
\begin{equation}
F=\sum_{h=0}^{N-1}a(h)T_{x}^{h}Y_{0}%
\end{equation}
for $a(h)$ a scaler. \ For the product we obtain
\begin{equation}
\sigma_{G}F=\sum_{r=0}^{N-1}\left(  \sum_{h=0}^{N-1}(h-r)a(h)\right)
T_{x}^{r}Y_{0}%
\end{equation}
and the value of the function $\sigma_{G}F$ at $T_{x}^{r}$ is the weighted sum
of the displacements from $T_{x}^{r}$ to the positions composing $F.$ \ For
$F$ normalized such that $\sum_{h=0}^{N-1}a(h)=1,$ we may interpret
$\sigma_{G}F$ evaluated at $T_{x}^{r}$ as the average displacement from
$T_{x}^{r}$ to $F$.

In general one may form test functions having a dependence on both $T_{x}^{r}
$ and $T_{y}^{s}$
\[
\sigma_{G}=\sum_{r,s=0}^{N-1}\sigma(-r,-s)T_{x}^{r}T_{y}^{s}%
\]
and consider normalized functions $F$
\[
F=\sum_{h,k=0}^{N-1}F(h,k)T_{x}^{h}T_{y}^{k}%
\]
The convolution%
\[
\sigma_{G}F=\sum_{r,s=0}^{N-1}\left(  \sum_{h,k=0}^{N-1}\sigma
(h-r,k-s)F(h,k)\right)  T_{x}^{r}T_{y}^{s}%
\]
evaluated at the origin $T_{x}^{0}T_{y}^{0}$ returns the value of
$\sigma(h,k)$ averaged over $F$. \ 

\subsubsection{\label{evaluation of test functions for P+fxf}Evaluation of
test functions in the system $P_{+}\ f\otimes f$}

Let us now consider some specific examples. \ Consider the convolution of the
test function
\begin{align}
\sigma_{G}  &  =X_{q}Y_{p}\\
&  =\sum_{m,n=0}^{N-1}\exp(\frac{-2\pi i}{N}(mq+np)\ T_{x}^{m}T_{y}%
^{n}\nonumber
\end{align}
with $P_{+}f\otimes f.$ \ Forming the product $\sigma_{G}P_{+}f\otimes f$ and
evaluating the result at the origin $T_{x}^{0}T_{y}^{0}$, we obtain from the
Fourier-Wigner form of $\ P_{+}f\otimes f$ (equation \ref{fwig})
\begin{gather}
\sigma_{G}P_{+}f\otimes f\mid_{0,0}=\nonumber\\
P_{+}\left(  \sum_{k=0}^{N-1}f(k)^{\ast}\exp\left(  \frac{2\pi i}%
{N}(qk+p\left(  \frac{-1}{2\omega}\frac{N}{2\pi i}\frac{d}{dk}\right)
\right)  f(k)\right)  \text{ }\label{fwigop}\\
\cdot\overline{X}_{0}\left(  I\otimes I\right) \nonumber
\end{gather}
From the Wigner distribution form equation \ref{wigd} we obtain
\begin{gather}
\sigma_{G}P_{+}f\otimes f\mid_{0,0}=\nonumber\\
\text{ }P_{+}\left(  \sum_{r.s,p=0}^{N-1}\exp(\frac{2\pi i}{N}%
(rq+sp))f(r+\frac{p}{4\omega})^{\ast}f(r-\frac{p}{4\omega})\exp(\frac{-2\pi
i}{N}ps)\right) \label{wigfunc}\\
\cdot\overline{X}_{0}\left(  I\otimes I\right)  /N\nonumber
\end{gather}
Comparing these two expressions we see that the expectation value of the
operator
\[
\exp\left(  \frac{2\pi i}{N}(qk+p\left(  \frac{-1}{2\omega}\frac{N}{2\pi
i}\frac{d}{dk}\right)  \right)
\]
in equation \ref{fwigop} is associated with the value of the function
\[
\exp(\frac{2\pi i}{N}(rq+sp))
\]
in equation \ref{wigfunc} evaluated over the Wigner distribution. \ This
particular correspondence between operator and function in the continuum case
is derived from the Weyl map (Weyl 1950 p275, Folland 1989 p81, Wong 1998
p21). \ This correspondence is rigorous in the present construction.

We next consider the case when $\sigma_{G}$ is a polynomial in $r$ and $s$
(cf., Folland 1989 p82). \ Expand the exponential in the expression for
$P_{+}f\otimes f$ \ equation \ref{fwig} in a power series.
\begin{gather}
P_{+}f\otimes f=\text{ }\label{fwigps}\\
P_{+}{\Huge \{}\sum_{k,p,q=0,h=0}^{N-1,\infty}f(k)^{\ast}\frac{1}{h!}\left(
\frac{2\pi i}{N}(qk-\frac{p}{2\omega}\left(  \frac{N}{2\pi i}\frac{d}%
{dk}\right)  \right)  ^{h}f(k)\text{ \ }X_{q}Y_{p}{\Huge \}}\nonumber\\
\text{\ \ \ \ \ \ \ }\cdot\text{\ }\overline{X}_{0}\left(  I\otimes I\right)
/N^{2}\nonumber
\end{gather}
The $h^{th}$ order of this expansion contains terms \
\begin{gather}
P_{+}\sum_{k=0}^{N-1}f(k)^{\ast}\left\{  \sum_{\text{all orderings}}%
k^{j}\left(  \frac{N}{2\pi i}\frac{d}{dk}\right)  ^{h-j}\right\}
f(k)\label{fwigpstm}\\
\cdot\text{ }\frac{1}{h!}\left(  \frac{2\pi i}{N}\right)  ^{h}\left(
\frac{-1}{2\omega}\right)  ^{h-j}\left(  \sum_{p,q=0}^{N-1}q^{j}%
p^{h-j}\text{\ }X_{q}Y_{p}\right) \nonumber\\
\text{\ }\cdot\text{\ }\overline{X}_{0}\left(  I\otimes I\right)
/N^{2}\nonumber
\end{gather}
where the bracketed term is a sum over all orderings of a product of $j$
factors of $k$ and $h-j$ factors of $\left(  \frac{N}{2\pi i}\frac{d}%
{dk}\right)  .$ \ Act on $P_{+}f\otimes f$ equation \ref{fwig} with the test
function
\begin{equation}
\sigma_{G}=\sum_{r,s=0}^{N-1}\left(  -r\right)  ^{m}(-s)^{n}T_{x}^{r}T_{y}^{s}%
\end{equation}
and use equation \ref{test f id} to conclude
\begin{gather}
\sigma_{G}P_{+}f\otimes f\mid_{0,0}=\nonumber\\
=P_{+}\text{ }\sum_{k=0}^{N-1}f(k)^{\ast}\left\{  \frac{m!n!}{(m+n)!}%
\sum_{\text{all orderings}}k^{m}\left(  \frac{-1}{2\omega}\frac{N}{2\pi
i}\frac{d}{dk}\right)  ^{n}\right\}  f(k)\text{ }\label{expop}\\
\cdot\text{\ }\overline{X}_{0}\left(  I\otimes I\right) \nonumber
\end{gather}
Now act with $\sigma_{G}$ on the Wigner distribution form of $P_{+}f\otimes f$
equation \ref{wigd} to conclude%
\begin{gather}
\sigma_{G}P_{+}f\otimes f\mid_{0,0}=\label{exfun1}\\
=P_{+}\sum_{r,s,p=0}^{N-1}r^{m}s^{n}f(r+\frac{p}{4\omega})^{\ast}f(r-\frac
{p}{4\omega})\exp(\frac{-2\pi i}{N}ps)\nonumber\\
\cdot\overline{X}_{0}\left(  I\otimes I\right)  /N\nonumber
\end{gather}
The expectation value of the \textquotedblright Weyl-ordered\textquotedblright%
\ operator
\begin{equation}
\frac{m!n!}{(m+n)!}\left(  \sum_{\text{all orderings}}k^{m}\left(  \frac
{-1}{2\omega}\frac{N}{2\pi i}\frac{d}{dk}\right)  ^{n}\right)  \label{weylord}%
\end{equation}
in equation \ref{expop} is associated with the evaluation of the function
$r^{m}s^{n}$ over the Wigner distribution (equation \ref{exfun1}).

Based on these results one is led to identify the convolution procedure
described above with the quantum measurement process for our system.

\section{\label{transformation under sl2}Transformation of $P_{+}f\otimes f$
under $g\otimes g$ for $g\in SL(2,N)$}

In this section we consider the transformation of $P_{+}f\otimes f$ \ under
$g\otimes g$ for $g\in SL(2,\mathbb{F}_{N}).\ \ $We decompose $P_{+}f\otimes
f$ \ into subspaces that are invariant under these transformations. \ Since
the subgroup $\{g\otimes g$ for $g\in SL(2,\mathbb{F}_{N})\}$ does not commute
with the subgroup $\{T_{x}^{r}T_{y}^{s}=t_{x}^{r}t_{y}^{s}\otimes t_{x}%
^{r}t_{y}^{s}$ for $r,s\in\mathbb{F}_{N}\}$ the subspaces that we find, except
for the trivial case, are not stable under the measurement process described
above. \ They are, however, useful for the description of a system that is
evolving under a sequence of transformations that are contained in
$SL(2,\mathbb{F}_{N}).$

Let us rewrite equation \ref{fwig} as
\begin{gather}
P_{+}f\otimes f=P_{+}\sum_{k,q,p=0}^{N-1}f(k)^{\ast}\exp\left(  \frac{2\pi
i}{N}\left(
\begin{array}
[c]{cc}%
q & p
\end{array}
\right)  \left(
\begin{array}
[c]{c}%
k\\
\frac{-1}{2\omega}\frac{N}{2\pi i}\frac{d}{dk}%
\end{array}
\right)  \right)  f(k)\text{ \ }X_{q}Y_{p}\text{ \ }\nonumber\\
\cdot\overline{X}_{0}\left(  I\otimes I\right)  /N^{2} \label{fwigmat}%
\end{gather}
where we view the argument of the exponential as arising from the contraction
of the matrix $\left(
\begin{array}
[c]{cc}%
q & p
\end{array}
\right)  $ with the matrix $\left(
\begin{array}
[c]{c}%
k\\
\frac{-1}{2\omega}\frac{N}{2\pi i}\frac{d}{dk}%
\end{array}
\right)  .$ \ In Appendix B we determine the action of $g\otimes g$ for $g\in
SL(2,N)$ on $f\otimes f.$ \ We find that the operation $g\otimes g$ can be
implemented in the following way
\begin{gather}
\left(  g\otimes g\right)  P_{+}f\otimes f=\nonumber\\
P_{+}\sum_{k,q,p=0}^{N-1}f(k)^{\ast}\exp\left(  \frac{2\pi i}{N}\left(
\begin{array}
[c]{cc}%
q & p
\end{array}
\right)  M_{g\otimes g}\left(
\begin{array}
[c]{c}%
k\\
\frac{-1}{2\omega}\frac{N}{2\pi i}\frac{d}{dk}%
\end{array}
\right)  \right)  f(k)\text{ \ }X_{q}Y_{p}\nonumber\\
\overline{\cdot X}_{0}\left(  I\otimes I\right)  /N^{2} \label{fwigmatg}%
\end{gather}
\qquad where $M_{g\otimes g}$ is a $2x2$ matrix . \ In particular we find
\begin{align}
M_{t_{s}^{a}\otimes t_{s}^{a}}  &  =\left(
\begin{array}
[c]{cc}%
1/a & 0\\
0 & a
\end{array}
\right) \\
M_{t_{u}^{b}\otimes t_{u}^{b}}  &  =\left(
\begin{array}
[c]{cc}%
1 & 0\\
-b & 1
\end{array}
\right) \\
M_{J\otimes J}  &  =\left(
\begin{array}
[c]{cc}%
0 & 1\\
-1 & 0
\end{array}
\right)
\end{align}

From a consideration of the power series expansion of equation \ref{fwigmatg}
we conclude that the subspace containing the $h^{th}$ order term is invariant
under $g\otimes g$ for $g\in SL(2,\mathbb{F}_{N})$ for each order $h$. \ The
$h^{th}$ order portion of $P_{+}f\otimes f$ is evident in equation
\ref{fwigps} and contains terms equation \ref{fwigpstm}. Let
\begin{gather}
\widehat{e}_{j,h-j}=P_{+}\frac{1}{j!(h-j)!}\left(  \frac{2\pi i}{N}\right)
^{h}\sum_{q,p=0}^{N-1}q^{j}p^{h-j}\text{\ }X_{q}Y_{p}\text{\ \ \ }\\
\cdot\text{\ }\overline{X}_{0}\left(  I\otimes I\right)  /N^{2}\nonumber
\end{gather}
and let
\begin{multline}
\left\langle k^{j}\left(  \frac{-1}{2\omega}\frac{N}{2\pi i}\frac{d}%
{dk}\right)  ^{h-j}\right\rangle =\nonumber\\
=\sum_{k=0}^{N-1}f(k)^{\ast}\left\{  \frac{j!(h-j)!}{h!}\sum_{\text{all
orderings}}k^{j}\left(  \frac{-1}{2\omega}\frac{N}{2\pi i}\frac{d}{dk}\right)
^{h-j}\right\}  f(k) \label{opdef}%
\end{multline}
We may then write the $h^{th}$ order term $r_{h}$ in the expansion of
$\ P_{+}f\otimes f$ as
\begin{equation}
r_{h}=\sum_{j=0}^{h}\left\langle k^{j}\left(  \frac{-1}{2\omega}\frac{N}{2\pi
i}\frac{d}{dk}\right)  ^{h-j}\right\rangle \widehat{e}_{j,h-j}%
\end{equation}
The basis vectors $\{\widehat{e}_{j,h-j}\mid0\leq j\leq h\}$ span an $h+1$
dimensional subspace.

The action of $g\otimes g$ on $r_{h}$ can be obtained from the $h^{th}$ order
term in the power series expansion of $\left(  g\otimes g\right)
P_{+}f\otimes f$. \ For a given one parameter family of transformations
$g\otimes g$ for $g\in SL(2,\mathbb{F}_{N})$ acting on $P_{+}f\otimes f$ one
may determine the evolution of $r_{h}$.

\subsection{Explicit realization of low-order subspaces of $P_{+}\left(
f\otimes f\right)  $}

\subsubsection{\label{zero order}Zeroth order}

Let $r_{0}$ denote the zero order term in the expansion $P_{+}f\otimes f.$
\ We have
\begin{equation}
r_{0}=\underset{k=0}{\sum^{N-1}}f(k)^{\ast}f(k)\widehat{e}_{0,0}%
\end{equation}
The action of the generators of $SL(2,N)$ on $r_{0}$ corresponds to the
trivial representation $\rho_{0}$:
\begin{equation}
\rho_{0}(J\otimes J)=\rho_{0}(t_{u}^{b}\otimes t_{u}^{b})=\rho_{0}(t_{s}%
^{a}\otimes t_{s}^{a})=1.
\end{equation}

\subsubsection{First order}

Let $r_{1}$ denote the first order term in the expansion of $P_{+}f\otimes f.
$ \ We have
\begin{equation}
r_{1}=\left\langle k\right\rangle \widehat{e}_{1,0}+\left\langle \frac
{-1}{2\omega}\frac{N}{2\pi i}\frac{d}{dk}\right\rangle \widehat{e}_{0,1}%
\end{equation}
Writing
\begin{equation}
r_{1}=\left(
\begin{array}
[c]{c}%
\left\langle k\right\rangle \\
\left\langle \frac{-1}{2\omega}\frac{N}{2\pi i}\frac{d}{dk}\right\rangle
\end{array}
\right)  \label{fstord}%
\end{equation}
the representations $\rho_{1}(g\otimes g)$ for the generators $g\in SL(2,N)$
acting on $r_{1}$ are given by the matrices
\begin{align}
\rho_{1}(t_{s}^{a}\otimes t_{s}^{a})  &  =\left(
\begin{array}
[c]{cc}%
1/a & 0\\
0 & a
\end{array}
\right)  ,\text{ }\rho_{1}(t_{u}^{b}\otimes t_{u}^{b})=\left(
\begin{array}
[c]{cc}%
1 & 0\\
-b & 1
\end{array}
\right) \nonumber\\
\text{ }\rho_{1}(J\otimes J)  &  =\left(
\begin{array}
[c]{cc}%
0 & 1\\
-1 & 0
\end{array}
\right)  ,\rho_{1}(t_{d}^{c}\otimes t_{d}^{c})=\left(
\begin{array}
[c]{cc}%
1 & -c\\
0 & 1
\end{array}
\right)
\end{align}
This representation of $g\otimes g$ for $g\in SL(2,\mathbb{F}_{N})$ is itself
equivalent to the standard representation of $SL(2,\mathbb{F}_{N})$ equation
\ref{sl2std}$.$ \ The bilinear form
\begin{equation}
b_{1}(r_{1}^{\prime},r_{1})=\left(  r_{1}^{\prime}\right)  ^{t}\left(
\begin{array}
[c]{cc}%
0 & 1\\
-1 & 0
\end{array}
\right)  r_{1},
\end{equation}
where $r_{1}$ and $r_{1}^{\prime}$ denote two vectors in the subspace spanned
by $\{\widehat{e}_{1,0},\widehat{e}_{0,1}\}$ and the superscript $t$ denotes
matrix transposition, is invariant under these transformations
\begin{equation}
b_{1}(\rho_{1}(g\otimes g)r_{1}^{\prime},\rho_{1}(g\otimes g)r_{1}%
)=b(r_{1}^{\prime},r_{1})
\end{equation}
The bilinear form $b_{1}(r_{1}^{\prime},r_{1})$ is the standard symplectic form.

\subsubsection{Second Order}

Let $r_{2}$ denote the second order term in the expansion of $P_{+}f\otimes
f.$ \ We have
\begin{equation}
r_{2}=\left\langle k^{2}\right\rangle \widehat{e}_{2,0}+\left\langle k\left(
\frac{-1}{2\omega}\frac{N}{2\pi i}\frac{d}{dk}\right)  \right\rangle
\widehat{e}_{1,1}+\left\langle \left(  \frac{-1}{2\omega}\frac{N}{2\pi i}%
\frac{d}{dk}\right)  ^{2}\right\rangle \widehat{e}_{0,2}%
\end{equation}
Writing $r_{2}$ as a column matrix
\begin{equation}
r_{2}=\left(
\begin{array}
[c]{c}%
\left\langle k^{2}\right\rangle \\
\left\langle k\left(  \frac{-1}{2\omega}\frac{N}{2\pi i}\frac{d}{dk}\right)
\right\rangle \\
\left\langle \left(  \frac{-1}{2\omega}\frac{N}{2\pi i}\frac{d}{dk}\right)
^{2}\right\rangle
\end{array}
\right)  \text{.}%
\end{equation}
the representations $\rho_{2}(g\otimes g)$ for the generators $g\in
SL(2,\mathbb{F}_{N})$ acting on $r_{2}$ are
\begin{align}
\rho_{2}(t_{s}^{a}\otimes t_{s}^{a})  &  =\left(
\begin{array}
[c]{ccc}%
1/a^{2} & 0 & 0\\
0 & 1 & 0\\
0 & 0 & a^{2}%
\end{array}
\right)  ,\text{ }\rho_{2}(t_{u}^{b}\otimes t_{u}^{b})=\left(
\begin{array}
[c]{ccc}%
1 & 0 & 0\\
-b & 1 & 0\\
b^{2} & -2b & 1
\end{array}
\right)  ,\\
\rho_{2}(J\otimes J)  &  =\left(
\begin{array}
[c]{ccc}%
0 & 0 & 1\\
0 & -1 & 0\\
1 & 0 & 0
\end{array}
\right)  ,\rho_{2}(t_{d}^{c}\otimes t_{d}^{c})=\left(
\begin{array}
[c]{ccc}%
1 & -2c & c^{2}\\
0 & 1 & -c\\
0 & 0 & 1
\end{array}
\right)  .\nonumber
\end{align}
We note that $\rho_{2}(J^{2}\otimes J^{2})=1$ so that $\rho_{2}$ is a
representation of $PSL(2,\mathbb{F}_{N})=SL(2,\mathbb{F}_{N})/\{1,J^{2}\}\sim
SO(1,2).$

The bilinear form
\begin{equation}
b_{2}(r_{2}^{\prime},r_{2})=\left(  r_{2}^{\prime}\right)  ^{t}\left(
\begin{array}
[c]{ccc}%
0 & 0 & 1\\
0 & -2 & 0\\
1 & 0 & 0
\end{array}
\right)  r_{2}%
\end{equation}
is invariant under $SL(2,\mathbb{F}_{N})$ transformations. \ For
$r_{2}^{\prime}=r_{2}$ we find the invariant bilinear form
\begin{equation}
\left\langle k^{2}\right\rangle \left\langle \left(  \frac{-1}{2\omega}%
\frac{N}{2\pi i}\frac{d}{dk}\right)  ^{2}\right\rangle -\left\langle k\left(
\frac{-1}{2\omega}\frac{N}{2\pi i}\frac{d}{dk}\right)  \right\rangle ^{2}%
\end{equation}
This invariant corresponds to the invariant $\left\langle \widehat{p}%
^{2}\right\rangle \left\langle \widehat{x}^{2}\right\rangle -\frac{1}%
{4}\left\langle \widehat{p}\widehat{x}+\widehat{x}\widehat{p}\right\rangle
^{2}$ that Dodonov and Man'ko describe (reviewed in Dodonov 2000 and Dodonov
and Man'ko 2000) for a 1d quantum system evolving under a homogeneous
quadratic Hamiltonian.

\subsubsection{Third Order}

Writing%
\begin{equation}
r_{3}=\left\langle \text{ }k^{3}\right\rangle \widehat{e}_{3,0}+\left\langle
k^{2}\left(  \frac{-1}{2\omega}\frac{N}{2\pi i}\frac{d}{dk}\right)
\right\rangle \widehat{e}_{2,1}+\left\langle k\left(  \frac{-1}{2\omega}%
\frac{N}{2\pi i}\frac{d}{dk}\right)  ^{2}\right\rangle \widehat{e}%
_{1,2}+\left\langle \left(  \frac{-1}{2\omega}\frac{N}{2\pi i}\frac
{d}{dk^{\prime}}\right)  ^{3}\right\rangle \widehat{e}_{0,3}%
\end{equation}
as a column vector we have the representation%
\begin{align}
\rho_{3}(t_{s}^{a}\otimes t_{s}^{a})  &  =\left(
\begin{array}
[c]{cccc}%
1/a^{3} & 0 & 0 & 0\\
0 & 1/a & 0 & 0\\
0 & 0 & a & 0\\
0 & 0 & 0 & a^{3}%
\end{array}
\right)  ,\text{ }\rho_{3}(t_{u}^{b}\otimes t_{u}^{b})=\left(
\begin{array}
[c]{cccc}%
1 & 0 & 0 & 0\\
-b & 1 & 0 & 0\\
b^{2} & -2b & 1 & 0\\
-b^{3} & 3b^{2} & -3b & 1
\end{array}
\right)  ,\nonumber\\
\rho_{3}(J\otimes J)  &  =\left(
\begin{array}
[c]{cccc}%
0 & 0 & 0 & 1\\
0 & 0 & -1 & 0\\
0 & 1 & 0 & 0\\
-1 & 0 & 0 & 0
\end{array}
\right)
\end{align}
and invariant bilinear form
\begin{equation}
b_{3}(r_{3}^{\prime},r_{3})=\left(  r_{3}^{\prime}\right)  ^{t}\left(
\begin{array}
[c]{cccc}%
0 & 0 & 0 & 1\\
0 & 0 & -3 & 0\\
0 & 3 & 0 & 0\\
-1 & 0 & 0 & 0
\end{array}
\right)  r_{3}%
\end{equation}

\subsubsection{\label{fourth order}Fourth Order}

Writing%

\begin{align}
r_{4}  &  =\left\langle \text{ }k^{4}\right\rangle \widehat{e}_{4,0}%
+\left\langle k^{3}\left(  \frac{-1}{2\omega}\frac{N}{2\pi i}\frac{d}%
{dk}\right)  \right\rangle \widehat{e}_{3,1}+\left\langle k^{2}\left(
\frac{-1}{2\omega}\frac{N}{2\pi i}\frac{d}{dk}\right)  ^{2}\right\rangle
\widehat{e}_{2,2}+\nonumber\\
&  +\left\langle k\left(  \frac{-1}{2\omega}\frac{N}{2\pi i}\frac{d}%
{dk}\right)  ^{3}\right\rangle \widehat{e}_{1,3}+\left\langle \left(
\frac{-1}{2\omega}\frac{N}{2\pi i}\frac{d}{dk^{\prime}}\right)  ^{4}%
\right\rangle \widehat{e}_{0,4}%
\end{align}
as a column vector we have the representation
\begin{align}
\rho_{4}(t_{s}^{a}\otimes t_{s}^{a})  &  =\left(
\begin{array}
[c]{ccccc}%
1/a^{4} & 0 & 0 & 0 & 0\\
0 & 1/a^{2} & 0 & 0 & 0\\
0 & 0 & 1 & 0 & 0\\
0 & 0 & 0 & a^{2} & 0\\
0 & 0 & 0 & 0 & a^{4}%
\end{array}
\right)  ,\text{ }\rho_{4}(t_{u}^{b}\otimes t_{u}^{b})=\left(
\begin{array}
[c]{ccccc}%
1 & 0 & 0 & 0 & 0\\
-b & 1 & 0 & 0 & 0\\
b^{2} & -2r & 1 & 0 & 0\\
-b^{3} & 3r^{2} & -3r & 1 & 0\\
b^{4} & -4r^{3} & 6r^{2} & -4r & 1
\end{array}
\right)  ,\nonumber\\
\text{ }\rho_{4}(J\otimes J)  &  =\left(
\begin{array}
[c]{ccccc}%
0 & 0 & 0 & 0 & 1\\
0 & 0 & 0 & -1 & 0\\
0 & 0 & 1 & 0 & 0\\
0 & -1 & 0 & 0 & 0\\
1 & 0 & 0 & 0 & 0
\end{array}
\right)
\end{align}
and invariant bilinear form
\begin{equation}
b_{4}(r_{4}^{\prime},r_{4})=\left(  r_{4}^{\prime}\right)  ^{t}\left(
\begin{array}
[c]{ccccc}%
0 & 0 & 0 & 0 & 1\\
0 & 0 & 0 & -4 & 0\\
0 & 0 & 6 & 0 & 0\\
0 & -4 & 0 & 0 & 0\\
1 & 0 & 0 & 0 & 0
\end{array}
\right)  r_{4}%
\end{equation}
For $r_{4}=r_{4}^{\prime}$ we obtain the invariant
\begin{equation}
2\left\langle \text{ }k^{4}\right\rangle \left\langle \left(  \frac
{-1}{2\omega}\frac{N}{2\pi i}\frac{d}{dk}\right)  ^{4}\right\rangle
-8\left\langle k\left(  \frac{-1}{2\omega}\frac{N}{2\pi i}\frac{d}{dk}\right)
^{3}\right\rangle \left\langle k^{3}\left(  \frac{-1}{2\omega}\frac{N}{2\pi
i}\frac{d}{dk}\right)  \right\rangle +6\left\langle k^{2}\left(  \frac
{-1}{2\omega}\frac{N}{2\pi i}\frac{d}{dk}\right)  ^{2}\right\rangle ^{2}%
\end{equation}
that corresponds to the invariant
\begin{equation}
\left\langle \widehat{p}^{4}\right\rangle \left\langle \widehat{x}%
^{4}\right\rangle +\frac{3}{4}\left(  \left\langle \widehat{p}^{2}\widehat
{x}^{2}+\widehat{x}^{2}\widehat{p}^{2}\right\rangle \right)  ^{2}-\frac{3}%
{2}\hbar^{2}\left\langle \widehat{p}^{2}\widehat{x}^{2}+\widehat{x}%
^{2}\widehat{p}^{2}\right\rangle -\left\langle \widehat{p}^{3}\widehat
{x}+\widehat{x}\widehat{p}^{3}\right\rangle \left\langle \widehat{p}%
\widehat{x}^{3}+\widehat{x}^{3}\widehat{p}\right\rangle
\end{equation}
described by Dodonov and Man'ko (Dodonov 2000 and Dodonov and Man'ko 2000).

\subsubsection{Higher order}

One may continue and obtain higher order representations of $SL(2,\mathbb{F}%
_{N})$ acting on $r_{h}$ for $h$ odd and of$\ PSL(2,\mathbb{F}_{N})$ for $h$
even (cf., e.g., Knapp 1986 p38, Fulton and Harris 1991 p150, and Howe and Tan
1992 pgs. 25, 55).

\subsection{\label{examples}2 examples}

Let us now consider the dynamics resulting from 2 familiar families of
1-parameter transformations $g\otimes g$ for $g\in SL(2,\mathbb{F}_{N})$.

\subsubsection{Free Particle}

Consider the 1-parameter family of transformations determined by $t_{d}%
^{\frac{-t}{m}}$ for $t,m\in\mathbb{F}_{N}.$ \ Using equation \ref{tdf} for
$t_{d}^{\frac{-t}{m}}$and assigning $\hbar=\frac{-1}{2\omega}\frac{N}{2\pi} $
we obtain
\begin{equation}
t_{d}^{\frac{-t}{m}}f=\sum_{k}^{N-1}\exp(it\frac{\hbar}{2m}\frac{d^{2}}%
{dk^{2}})f(k)t_{x}^{k}I. \label{freep}%
\end{equation}
Differentiating with respect to $t$ by expanding out the exponential in
equation \ref{freep} and using equation \ref{deriv} results in the
Schr\"{o}dinger equation for a free particle (cf. de Gossen 2001 p273).

The first order subspace equation \ref{fstord} transforms with $t$ according
to
\begin{equation}
\rho_{1}(t_{d}^{\frac{-t}{m}}\otimes t_{d}^{\frac{-t}{m}})\left(
\begin{array}
[c]{c}%
\left\langle k\right\rangle \\
\left\langle \frac{\mathbb{\hbar}}{i}\frac{d}{dk}\right\rangle
\end{array}
\right)  =\left(
\begin{array}
[c]{c}%
\left\langle k\right\rangle +\frac{t}{m}\left\langle \frac{\mathbb{\hbar}}%
{i}\frac{d}{dk}\right\rangle \\
\left\langle \frac{\mathbb{\hbar}}{i}\frac{d}{dk}\right\rangle
\end{array}
\right)
\end{equation}
corresponding to the motion of a free particle with constant momentum
$\left\langle \frac{\mathbb{\hbar}}{i}\frac{d}{dk}\right\rangle $ and velocity
$\left\langle \frac{\mathbb{\hbar}}{i}\frac{d}{dk}\right\rangle /m.$ \ One may
also readily calculate the evolution of the higher order subspaces.

\subsubsection{Simple Harmonic Oscillator}

We consider the 1-parameter family of transformations determined by%
\begin{equation}
t_{r}^{n}=\left(
\begin{array}
[c]{cc}%
a & b\delta\\
b & a
\end{array}
\right)  ^{n}%
\end{equation}
for $\delta$ a nonsquare in $\mathbb{F}_{N}$, $b\neq0$ and $0\leq n<N+1.$
\ Powers of this matrix can be calculated by mapping to the isomorphic
multiplicative group within the quadratic extension of $\mathbb{F}_{N}$
\begin{equation}
\left(
\begin{array}
[c]{cc}%
a & b\delta\\
b & a
\end{array}
\right)  \longrightarrow a+b\sqrt{\delta}%
\end{equation}
Writing $\left(  a+b\sqrt{\delta}\right)  ^{N+1}=\left(  a+b\sqrt{\delta
}\right)  \left(  a+b\sqrt{\delta}\right)  ^{N}$ and using the binomial
expansion of $\left(  a+b\sqrt{\delta}\right)  ^{N},$ where $\sqrt{\delta}%
^{N}=\sqrt{\delta}\delta^{\frac{N-1}{2}}=-\sqrt{\delta}$ , $a^{N}=a,$ and
$b^{N}=b,$ we find $\left(  a+b\sqrt{\delta}\right)  ^{N+1}=a^{2}-b^{2}%
\delta=1$ so that the group generated by $t_{r}$ has order $N+1$ (see, e.g.,
Terras 1999 p307)$.$ \ 

For the first order subspace equation \ref{fstord} we find the evolution with
$n$
\begin{gather}
\rho_{1}(t_{r}^{n}\otimes t_{r}^{n})\left(
\begin{array}
[c]{c}%
\left\langle k\right\rangle \\
\left\langle \frac{-1}{2\omega}\frac{N}{2\pi i}\frac{d}{dk}\right\rangle
\end{array}
\right)  =\left(
\begin{array}
[c]{cc}%
a & -b\\
-b\delta & a
\end{array}
\right)  ^{n}\left(
\begin{array}
[c]{c}%
\left\langle k\right\rangle \\
\left\langle \frac{-1}{2\omega}\frac{N}{2\pi i}\frac{d}{dk}\right\rangle
\end{array}
\right) \\
=\frac{1}{2}(\left\langle k\right\rangle +\frac{1}{\sqrt{\delta}}\left\langle
\frac{-1}{2\omega}\frac{N}{2\pi i}\frac{d}{dk}\right\rangle )\lambda_{+}%
^{n}\mathbf{e}_{+}+\frac{1}{2}(\left\langle k\right\rangle -\frac{1}%
{\sqrt{\delta}}\left\langle \frac{-1}{2\omega}\frac{N}{2\pi i}\frac{d}%
{dk}\right\rangle )\lambda_{-}^{n}\mathbf{e}_{-}\nonumber
\end{gather}
where $\lambda_{\pm}=a\mp b\sqrt{\delta\text{ }}$ are eigenvalues
corresponding to eigenvectors $\mathbf{e}_{\pm}=\left(
\begin{array}
[c]{c}%
1\\
\pm\sqrt{\delta}%
\end{array}
\right)  $ of $\rho_{1}(t_{r}\otimes t_{r}).$ \ We have $\lambda_{+}%
\lambda_{-}=1$ so that $\lambda_{-}^{n}=\lambda_{+}^{-n}.$ \ Letting
\begin{align}
c(n)  &  =\frac{1}{2}(\lambda_{+}^{n}+\lambda_{+}^{-n})\\
s(n)  &  =\frac{1}{2\sqrt{\delta}}(\lambda_{+}^{n}-\lambda_{+}^{-n})
\end{align}
we may write%
\begin{equation}
\rho_{1}(t_{r}^{n}\otimes t_{r}^{n})\left(
\begin{array}
[c]{c}%
\left\langle k\right\rangle \\
\left\langle \frac{-1}{2\omega}\frac{N}{2\pi i}\frac{d}{dk}\right\rangle
\end{array}
\right)  =\left(
\begin{array}
[c]{cc}%
c(n) & s(n)\\
s(n)\delta & c(n)
\end{array}
\right)  \left(
\begin{array}
[c]{c}%
\left\langle k\right\rangle \\
\left\langle \frac{-1}{2\omega}\frac{N}{2\pi i}\frac{d}{dk}\right\rangle
\end{array}
\right)
\end{equation}
One may similarly derive the evolution of the higher order subspaces in the
direct product algebra. \ 

Let us defer for now a more detailed treatment of this case. \ Athanasiu and
Floratos 1994 consider the case $\delta=-1.$ \ For $N=3\operatorname{mod}4,$
$\delta=-1$ is a nonsquare in $\mathbb{F}_{N}$ and so in this instance
corresponds to the case that we consider above. \ Additional related work for
the case $\delta=-1$ are contained in Balian and Itzykson 1986, Floratos and
Leontaris 1997,\ Athanasiu et.al. 1996 and Floratos and Nicolis 2005.

\section{Discussion}

The main idea guiding this paper is the hypothesis that physical observables
reside in a space formed by taking the tensor product of 2 copies of an
element in an invariant subspace of an underlying group algebra. \ The group
of transformations leaving this space invariant is identified with the allowed
transformations of the physical observables. \ A methodology for calculating
the expectation value for physical observables then follows. \ The
construction that I describe is generic. \ 

In this paper we have considered this construction for the case of the Jacobi
group that is the semidirect product of the Heisenberg group $H_{1}%
(\mathbb{F}_{N})$ with its automorphism group $SL(2,\mathbb{F}_{N})$. \ The
goal has been to perform this construction for the finite group that provides
a model counterpart for the continuum quantum case of a single particle with a
single spatial degree of freedom and whose phase space is a flat torus.

We have constructed the Schr\"{o}dinger-Weil representation of the Jacobi
group whose restriction to the Heisenberg subgroup corresponds to the
Schr\"{o}dinger representation and whose restriction to $SL(2,\mathbb{F}_{N})$
corresponds to the metaplectic representation. \ This representation is
concretely realized by the left action of the Jacobi group on functions
defined in a particular ideal of the Jacobi group algebra.

We have defined a derivative operation acting on a function contained in the
regular representation of a cyclic group. \ This derivative operation allows a
treatment for the prime field case that parallels the characteristic zero case.

We take an element $f$ that is contained within the chosen ideal of the Jacobi
group algebra and form the direct product $f\otimes f$. \ The Hermitian
portion $P_{+}f\otimes f$ resides in the group algebra of the group
$\{T_{x}^{r}\otimes T_{y}^{s}=t_{x}^{r}t_{y}^{s}\otimes t_{x}^{r}t_{y}^{s}\mid
r,s\in\mathbb{F}_{N}\}$; \ this portion corresponds to the Wigner distribution
of $f$. \ For normalized $P_{+}f\otimes f$, the convolution of $P_{+}f\otimes
f$ \ with a test function $\sigma_{G}=\sum_{r,s\in\mathbb{F}_{N}}%
\sigma(-r,-s)T_{x}^{r}T_{y}^{s}$ has value at the origin $T_{x}^{0}T_{y}^{0}$
equal to $\sigma(r,s)$ evaluated over the Wigner distribution of $f$. \ The
Weyl map that relates the value of $\sigma(r,s)$ evaluated over the Wigner
distribution with the expectation value of a particular operator is then
obtained. \ The operator expressions that we derive are in the Weyl ordered
form and there is no ordering ambiguity.

We next consider the transformation of \ $P_{+}f\otimes f$ under $g\otimes g$
for $g\in SL(2,\mathbb{F}_{N})$ and find invariant subspaces. \ These
subspaces correspond to representations of $SL(2,\mathbb{F}_{N})$ that can be
realized on spaces of homogeneous polynomials in 2 real variables. \ The
$n^{th}$ order subspace with dimension $n+1$ is associated with homogeneous
polynomials of degree $n.$ \ \ The zeroth order subspace is a constant under
$g\otimes g$ and after normalization can be interpreted as the total
probability of the quantum system. \ The first order subspace is 2-dimensional
with coordinates $\left\langle k\right\rangle $ and $\left\langle \frac
{-1}{2\omega}\frac{N}{2\pi i}\frac{d}{dk}\right\rangle $ and conserved
symplectic form. \ This subspace can be associated with the classical phase
space of the physical system. \ Higher order subspaces and their conserved
bilinear forms are described. \ The quadratic forms for the 2nd and 4th order
subspaces correspond to "quantum universal invariants" described by Dodonov
2000 and Dodonov and Man'ko 2000 for quantum systems evolving under a
homogeneous quadratic Hamiltonian.

We describe the 1 parameter families of transformations associated with the
motion of a free particle and with the simple harmonic oscillator.

We have considered only the Schr\"{o}dinger-Weil representation of the Jacobi
group. \ It is interesting to consider the nature of the physical systems
described by the other representations.

Let us now detail the correspondence between particular constructions in the
quantum mechanical treatment of a single particle with a single translational
degree of freedom and the construction developed in this paper:

\bigskip\bigskip%
\begin{tabular}
[c]{c|c}%
\textbf{Quantum Mechanics of particle } & \textbf{Construction }\\
\textbf{with 1 degree of freedom} & \textbf{of this paper}\\\hline
Configuration space & Invariant subspace\\
& of Jacobi group algebra\\\hline
Wavefunction & Element in invariant subspace\\
& of Jacobi group algebra\\\hline
Heisenberg constant & $\frac{-1}{2\omega}$ where $exp(\frac{2\pi i}{N}\omega)$
is character\\
& of $t_{z}$ $\in center$ of $G^{J}$\\\hline
Wigner function & $P_{+}f\otimes f$\\\hline
Weyl map & Obtained from convolution\\
& of $P_{+}f\otimes f$ with test function\\\hline
Expectation value & Obtained from convolution\\
of operator & with test function and\\
& evaluation at origin of algebra\\\hline
Total probability & Zero order subspace of $P_{+}f\otimes f$\\\hline
Classical phase space & First order subspace of $P_{+}f\otimes f$\\\hline
Higher order quantum & From bilinear forms for\\
invariants & higher order subspaces\\
& of $P_{+}f\otimes f$%
\end{tabular}

\appendix

\section{Action of J on ideal determined by $\widehat{y}_{0}\widehat
{z}_{\omega}\widehat{s}_{-}\widehat{u}_{0}\widehat{x}_{0}(1+\alpha
J)\widehat{d}_{0}$}

Consider the action of $J$ on an element $f$
\begin{equation}
f=\sum_{k=0}^{N-1}f(k)t_{x}^{k}\widehat{y}_{0}\widehat{z}_{\omega}\widehat
{s}_{-}\widehat{u}_{0}\widehat{x}_{0}(1+\alpha J)\widehat{d}_{0}%
\end{equation}
where $\alpha$ is a number that is to be determined. \ Let us consider the
action of $J$ on the first portion of this expression:
\begin{align}
J\sum_{k=0}^{N-1}f(k)t_{x}^{k}\widehat{y}_{0}\widehat{z}_{\omega}  &
=\sum_{k,l=0}^{N-1}f(k)Jt_{x}^{k}t_{y}^{l}\widehat{z}_{\omega}\\
&  =\sum_{k,l=0}^{N-1}f(k)t_{x}^{l}t_{y}^{-k}t_{z}^{2kl}\widehat{z}_{\omega}J
\end{align}
Inserting $1=\frac{1}{N}\sum_{\nu}\widehat{y}_{\upsilon}$ we obtain
\begin{equation}
=\frac{1}{N}\sum_{k,l,\nu=0}^{N-1}f(k)\exp(\frac{2\pi i}{N}(2kl\omega
-k\nu)t_{x}^{l}\widehat{y}_{\upsilon}\widehat{z}_{\omega}J
\end{equation}
Using the identity
\begin{equation}
t_{x}^{\nu/2\omega}\widehat{y}_{\upsilon}\widehat{z}_{\omega}=\widehat{y}%
_{0}t_{x}^{\nu/2\omega}\widehat{z}_{\omega}%
\end{equation}
we obtain%

\begin{equation}
=\frac{1}{N}\sum_{k,l,\nu=0}^{N-1}f(k)\exp(\frac{2\pi i}{N}(2\omega
k)(l-\frac{\nu}{2\omega})t_{x}^{l-\frac{\nu}{2\omega}}\widehat{y}_{0}%
t_{x}^{\nu/2\omega}\widehat{z}_{\omega}J
\end{equation}
Letting $l^{\prime}=l-\frac{\nu}{2\omega}$ and summing over $\nu$ \ we obtain
\begin{equation}
Jf=\frac{1}{N}\sum_{k,l^{\prime}}^{N-1}f(k)\exp(\frac{2\pi i}{N}2\omega
kl^{\prime})t_{x}^{l^{\prime}}\widehat{y}_{0}\widehat{z}_{\omega}%
\underline{\widehat{x}_{0}J}\widehat{s}_{-}\widehat{u}_{0}\widehat{x}%
_{0}(1+\alpha J)\widehat{d}_{0} \label{fststep}%
\end{equation}
We conclude that the action of $J$ on the first portion of $f$ leads to the
inverse Fourier transformation of $f(k)$ and we now have the term $\widehat
{x}_{0}J$ (underlined in equation \ref{fststep}) acting on the second part of
equation \ref{fststep}.\qquad

It is clearer to do the next part of this calculation in two parts. \ Let us
first consider the action of $\widehat{x}_{0}J$ on the term in equation
\ref{fststep} that contains $\alpha J$ and expand out $\widehat{u}_{0}%
\widehat{x}_{0}.$ \ We obtain
\begin{align}
\widehat{y}_{0}\widehat{z}_{\omega}\underline{\widehat{x}_{0}J}\widehat{s}%
_{-}\widehat{u}_{0}\widehat{x}_{0}J\widehat{d}_{0}  &  =\widehat{y}%
_{0}\widehat{z}_{\omega}\underline{\widehat{x}_{0}J}\widehat{s}_{-}\left(
\sum_{r,m=0}^{N-1}t_{u}^{m}t_{x}^{r}\right)  J\widehat{d}_{0}\nonumber\\
&  =\widehat{y}_{0}\widehat{z}_{\omega}\underline{\widehat{x}_{0}J}\widehat
{s}_{-}\sum_{r,m=0}^{N-1}t_{y}^{-ms}t_{x}^{r}t_{z}^{-mr^{2}}t_{u}^{m}%
J\widehat{d}_{0}%
\end{align}
Insert $1=JJ^{-1}$ to the left of $t_{u}^{m}.$ \ The resulting term
$J^{-1}t_{u}^{m}J$ can be absorbed\ into $\widehat{d}_{0}$ leaving
\begin{equation}
\widehat{y}_{0}\widehat{z}_{\omega}\underline{\widehat{x}_{0}J}\widehat{s}%
_{-}\widehat{u}_{0}\widehat{x}_{0}J\widehat{d}_{0}=\widehat{y}_{0}\widehat
{z}_{\omega}\underline{\widehat{x}_{0}J}\widehat{s}_{-}\left(  \sum
_{r,m=0}^{N-1}t_{y}^{-mr}t_{x}^{r}t_{z}^{-mr^{2}}\right)  J\widehat{d}_{0}%
\end{equation}
\ $t_{y}^{-mr}$ passes to the left to be absorbed into $\widehat{x}_{0}$ and
$t_{z}^{-mr^{2}}$ converts into the exponential $exp(\frac{-2\pi i}{N}%
mr^{2}\omega)$ after acting on $\widehat{z}_{\omega}.$ \ The sum over $m$\ of
this exponential has support only for $r=0.$ \ We are left with
\begin{align}
\widehat{y}_{0}\widehat{z}_{\omega}\underline{\widehat{x}_{0}J}\widehat{s}%
_{-}\widehat{u}_{0}\widehat{x}_{0}J\widehat{d}_{0}  &  =N\widehat{y}%
_{0}\widehat{z}_{\omega}\underline{\widehat{x}_{0}J^{2}}\widehat{s}%
_{-}\widehat{d}_{0}\label{step1}\\
&  =N\left(  \frac{-1}{N}\right)  \widehat{y}_{0}\widehat{z}_{\omega}%
\widehat{s}_{-}\widehat{x}_{0}\widehat{d}_{0}%
\end{align}
where
\begin{align}
J^{2}\widehat{s}_{-}  &  =t_{s}^{-1}\widehat{s}_{-}\nonumber\\
&  =\left(  \frac{-1}{N}\right)  \widehat{s}_{-}=\left\{
\begin{array}
[c]{c}%
+1\text{ for }N=1\text{ modulo }4\\
-1\text{ for }N=3\text{ modulo }4
\end{array}
\right\}  \widehat{s}_{-}%
\end{align}
\qquad

Now consider the term $\widehat{y}_{0}\widehat{z}_{\omega}\underline
{\widehat{x}_{0}J}\widehat{s}_{-}\widehat{u}_{0}\widehat{x}_{0}\widehat{d}_{0}
$ in equation \ref{fststep}. \ Commute $J$ through $\widehat{s}_{-}$ and
expand out the resulting term $J\widehat{u}_{0}$%
\begin{align}
\widehat{y}_{0}\widehat{z}_{\omega}\widehat{x}_{0}\widehat{s}_{-}J\widehat
{u}_{0}\widehat{x}_{0}\widehat{d}_{0}  &  =\widehat{y}_{0}\widehat{z}_{\omega
}\widehat{x}_{0}\widehat{s}_{-}\left(  \sum_{m=0}^{N-1}Jt_{u}^{m}\right)
\widehat{x}_{0}\widehat{d}_{0}\nonumber\\
&  =\widehat{y}_{0}\widehat{z}_{\omega}\widehat{x}_{0}\widehat{s}_{-}\left(
\sum_{m=1}^{N-1}t_{s}^{1/m}t_{u}^{-m}J^{-1}t_{u}^{-1/m}J^{-1}+J\right)
\widehat{x}_{0}\widehat{d}_{0}\nonumber\\
&  =\widehat{y}_{0}\widehat{z}_{\omega}\widehat{x}_{0}\widehat{s}_{-}\left(
\sum_{m=1}^{N-1}\left(  \frac{-m}{N}\right)  t_{u}^{-m}+J\right)  \widehat
{x}_{0}\widehat{d}_{0} \label{intermeq}%
\end{align}
where in the second line we use equation \ref{Jtu} to reexpress $Jt_{u}^{m}$
and in the third line $t_{s}^{1/m}$ converts into $\left(  \frac{1/m}%
{N}\right)  =\left(  \frac{m}{N}\right)  $ after acting on $\widehat{s}_{-}.$
\ In the second line we write $J^{-1}t_{u}^{-\frac{1}{m}}J^{-1}=t_{s}%
^{-1}Jt_{u}^{-\frac{1}{m}}J^{-1}$. \ \ $t_{s}^{-1}$ then changes the sign of
the Legendre symbol $\left(  \frac{m}{N}\right)  $ after acting on
$\widehat{s}_{-}$ and $Jt_{u}^{-\frac{1}{m}}J^{-1}$ is absorbed into
$\widehat{d}_{0}$ after commuting through $\widehat{x}_{0}.$

It is clearer now to consider separately the two terms in equation
\ref{intermeq}. \ For the first term, commute $\widehat{x}_{0}$ through
$\widehat{s}_{-}$ and expand $\widehat{x}_{0}$ to obtain
\begin{multline*}
\widehat{y}_{0}\widehat{z}_{\omega}\widehat{s}_{-}\left(  \sum_{m=1,h=0}%
^{N-1}\left(  \frac{-m}{N}\right)  t_{x}^{h}t_{u}^{-m}\right)  \widehat{x}%
_{0}\widehat{d}_{0}=\\
=\widehat{y}_{0}\widehat{z}_{\omega}\widehat{s}_{-}\left(  \sum_{m=1,h=0}%
^{N-1}\left(  \frac{-m}{N}\right)  t_{u}^{-m}t_{y}^{-mh}t_{x}^{h}%
t_{z}^{-mh^{2}}\right)  \widehat{x}_{0}\widehat{d}_{0}%
\end{multline*}
Now pass $t_{y}^{-mh}$ to the left to be absorbed into $\widehat{y}_{0},$ pass
$t_{x}^{h}$ to the right to be absorbed into $\widehat{x}_{0}$ and convert
$t_{z}^{-mh^{2}}$ into an exponential by acting on $\widehat{z}_{\omega}.$
\ Sum over $h$ in this exponential to obtain
\begin{equation}
\sum_{h=0}^{N-1}\exp(\frac{-2\pi i}{N}mh^{2}\omega)=\left(  \frac{-m}%
{N}\right)  \left(  \frac{\omega}{N}\right)  G(1,N)
\end{equation}
where $G(1,N)$ is the Gauss sum (see, e.g., Lang 1994 p85)
\begin{equation}
G(1,N)={\Huge \{}_{i\sqrt{N}for\text{ }N=3\operatorname{mod}4}^{\sqrt
{N}for\text{ }N=1\operatorname{mod}4}%
\end{equation}
\ Now combining these steps we have for the first term in equation
\ref{intermeq}
\begin{gather}
\widehat{y}_{0}\widehat{z}_{\omega}\widehat{x}_{0}\widehat{s}_{-}\left(
\sum_{m=1}^{N-1}\left(  \frac{-m}{N}\right)  t_{u}^{-m}\right)  \widehat
{x}_{0}\widehat{d}_{0}=\text{ \ \ \ \ \ \ \ \ \ \ \ \ \ \ }\nonumber\\
\text{ \ \ \ \ \ \ \ \ \ \ \ \ \ \ \ \ \ \ \ \ \ \ \ \ \ \ \ \ \ \ \ \ \ }%
=\widehat{y}_{0}\widehat{z}_{\omega}\widehat{s}_{-}\left(  \sum_{m=1}%
^{N-1}t_{u}^{-m}\left(  \frac{-m}{N}\right)  ^{2}\left(  \frac{\omega}%
{N}\right)  G(1,N)\right)  \widehat{x}_{0}\widehat{d}_{0}\nonumber\\
\text{ \ \ \ \ \ \ \ \ \ \ \ \ \ \ }=\widehat{y}_{0}\widehat{z}_{\omega
}\widehat{s}_{-}\left(  \frac{\omega}{N}\right)  G(1,N)(\widehat{u}%
_{0}-1)\widehat{x}_{0}\widehat{d}_{0} \label{step2}%
\end{gather}
where in the second line we use $\left(  \frac{-m}{N}\right)  ^{2}=1$ and sum
over $m$ to obtain the third line$.$

Let us now consider the second term in equation \ref{intermeq}. \ \ We commute
$\widehat{x}_{0}$ to the right through $\widehat{s}_{-}$ and use $J\widehat
{x}_{0}=\widehat{y}_{0}J$ to obtain
\begin{align}
\widehat{y}_{0}\widehat{z}_{\omega}\widehat{x}_{0}\widehat{s}_{-}J\widehat
{x}_{0}\widehat{d}_{0}  &  =\widehat{y}_{0}\widehat{z}_{\omega}\widehat{s}%
_{-}\widehat{x}_{0}\widehat{y}_{0}J\widehat{d}_{0}\nonumber\\
&  =\widehat{y}_{0}\widehat{z}_{\omega}\widehat{s}_{-}\sum_{h,l=0}^{N-1}%
t_{x}^{h}t_{y}^{l}J\widehat{d}_{0}\nonumber\\
&  =\widehat{y}_{0}\widehat{z}_{\omega}\widehat{s}_{-}\sum_{h,l=0}^{N-1}%
t_{y}^{l}t_{x}^{h}t_{z}^{2hl}J\widehat{d}_{0}%
\end{align}
$t_{y}^{l}$ passes to the left to be absorbed into $\widehat{y}_{0}$ and
$t_{z}^{2hl}$ converts to the exponential $\exp(\frac{2\pi i}{N}2hl\omega).$
\ The sum over $l$ of this exponential has support only for $h=0.$ \ We are
left with
\begin{equation}
\widehat{y}_{0}\widehat{z}_{\omega}\widehat{x}_{0}\widehat{s}_{-}J\widehat
{x}_{0}\widehat{d}_{0}=\widehat{y}_{0}\widehat{z}_{\omega}\widehat{s}%
_{-}NJ\widehat{d}_{0} \label{step3}%
\end{equation}

Combining now the results of equations \ref{step1}, \ref{step2} and
\ref{step3} we have
\begin{align}
Jf  &  =J\sum_{k=0}^{N-1}f(k)t_{x}^{k}\widehat{y}_{0}\widehat{z}_{\omega
}\widehat{s}_{-}\widehat{u}_{0}\widehat{x}_{0}(1+\alpha J)\widehat{d}%
_{0}\nonumber\\
&  =\frac{1}{N}\sum_{k,l=0}^{N-1}f(k)\exp(\frac{2\pi i}{N}2kl\omega)t_{x}%
^{l}\widehat{y}_{0}\widehat{z}_{\omega}\widehat{s}_{-}\nonumber\\
&  \cdot\left\{  \left(  \frac{\omega}{N}\right)  G(1,N)(\widehat{u}%
_{0}-1)\widehat{x}_{0}+NJ+\alpha N\left(  \frac{-1}{N}\right)  \widehat{x}%
_{0}\right\}  \widehat{d}_{0}%
\end{align}
Using the identity $\widehat{y}_{0}\widehat{z}_{\omega}\widehat{s}%
_{-}NJ\widehat{d}_{0}=\widehat{y}_{0}\widehat{z}_{\omega}\widehat{s}%
_{-}\widehat{u}_{0}\widehat{x}_{0}J\widehat{d}_{0}$ to reexpress the $NJ$ term
in the above expression we have
\begin{multline}
Jf=\frac{\left(  \frac{\omega}{N}\right)  G(1,N)}{N}\sum_{k,l=0}^{N-1}%
f(k)\exp(\frac{2\pi i}{N}kl\omega)t_{x}^{l}\widehat{y}_{0}\widehat{z}_{\omega
}\widehat{s}_{-}\\
\cdot\left[  \widehat{u}_{0}\widehat{x}_{0}(1+\frac{1}{\left(  \frac{\omega
}{N}\right)  G(1,N)}J)-(1-\frac{\alpha N\left(  \frac{-1}{N}\right)  }{\left(
\frac{\omega}{N}\right)  G(1,N)})\widehat{x}_{0}\right]  \widehat{d}%
_{0}\nonumber
\end{multline}
For
\[
\alpha=\frac{1}{\left(  \frac{\omega}{N}\right)  G(1,N)}%
\]
the second term in the above bracket vanishes and we have our result
\begin{multline}
Jf=\nonumber\\
=\frac{\left(  \frac{\omega}{N}\right)  G(1,N)}{N}\sum_{k,l=0}^{N-1}%
f(k)\exp(\frac{2\pi i}{N}2\omega kl)t_{x}^{l}\widehat{y}_{0}\widehat
{z}_{\omega}\widehat{s}_{-}\widehat{u}_{0}\widehat{x}_{0}\left[  1+\frac
{1}{\left(  \frac{\omega}{N}\right)  G(1,N)}J\right]  \widehat{d}_{0}%
\end{multline}

\section{Action of $g\otimes g$ on $f\otimes f$}

We now consider the transformation of $f\otimes f$ under the action of
$g\otimes g$ for
\[
f=\sum_{k=0}^{N-1}f(k)t_{x}^{k}I
\]%
\[
I=\widehat{y}_{0}\widehat{z}_{\omega}\widehat{s}_{-}\widehat{u}_{0}\widehat
{x}_{0}\left(  1+\frac{1}{\left(  \frac{\omega}{N}\right)  G(1,N)}J\right)
\widehat{d}_{0}%
\]
and $g\in SL(2,\mathbb{F}_{N}).$ \ Below I show the calculation for $J\otimes
J$ in detail. \ The calculations for $t_{s}^{a}\otimes t_{s}^{a}$ and
$t_{u}^{b}\otimes t_{u}^{b}$ are very similar and I merely give the final result.

Consider the action of $J\otimes J$ on $f\otimes f:$ \ Using equation
\ref{JxJ} we have\
\begin{gather}
\left(  J\otimes J\right)  f\otimes f=\nonumber\\
\kappa\left(  \sum_{k,k^{\prime},h,h^{\prime}=0}^{N-1}f(k)\exp\left(
\frac{2\pi i}{N}(2\omega kk^{\prime})\right)  t_{x}^{k^{\prime}}\otimes
f(h)\exp\left(  \frac{2\pi i}{N}(2\omega hh^{\prime})\right)  t_{x}%
^{h^{\prime}}\right) \nonumber\\
\cdot\left(  I\otimes I\right)  /N
\end{gather}
where
\begin{equation}
\kappa={\Huge \{}_{E=i\otimes i\text{ for }N=3\operatorname{mod}4}^{1\text{
for }N=1\operatorname{mod}4}%
\end{equation}
Changing variables using $p=h-k$, $p^{\prime}=h^{\prime}-k^{\prime},$ using
$T_{x}^{p^{\prime}/2}\overline{T}_{x}^{p^{\prime}/2}=1\otimes t_{x}%
^{p^{\prime}}$ and inserting $1=\frac{1}{N}\sum_{q=0}^{N-1}X_{q}$ we obtain
\begin{gather}
Jf\otimes Jf=\kappa\sum_{k,k^{\prime},l,l^{\prime}=0}^{N-1}f(k)\otimes
f(k+p)\nonumber\\
\cdot\exp\left(  \frac{2\pi i}{N}(-Ekk^{\prime}2\omega+(p+k)(p^{\prime
}+k^{\prime})2\omega+q(k^{\prime}+\frac{p^{\prime}}{2})\right) \nonumber\\
\cdot X_{q}\overline{T}_{x}^{p^{\prime}/2}\left(  I\otimes I\right)  /N^{2}%
\end{gather}
We now sum over $k^{\prime}$ to obtain $p=\frac{-q}{2\omega}-k(1-E).$
\ Substituting in for $p$ and reexpressing the exponential we obtain
\begin{gather}
Jf\otimes Jf=\text{ \ \ }{\Huge [}P_{+}\sum_{k,l,p=0}^{N-1}f(k)^{\ast}%
\exp\left(  \frac{2\pi i}{N}(p^{\prime}2\omega k-\frac{q}{2\omega}(\frac
{N}{2\pi i}\frac{d}{dk}))\right)  f(k)\\
+\kappa P_{-}\sum_{k,l,p=0}^{N-1}f(k)\exp\left(  \frac{-2\pi i}{N}(p^{\prime
}2\omega k-\frac{q}{2\omega}(\frac{N}{2\pi i}\frac{d}{dk}))\right)
f(-k){\Huge ]}\nonumber\\
\cdot\text{\ }X_{q}\overline{T}_{x}^{p^{\prime}/2}\left(  I\otimes I\right)
/N\nonumber
\end{gather}

Let us rewrite equation \ref{fxf}
\begin{gather}
f\otimes f=\sum_{k,l,p=0}^{N-1}f(k)\otimes\exp\left(  \frac{2\pi i}{N}\left(
\begin{array}
[c]{cc}%
q & p
\end{array}
\right)  \left(
\begin{array}
[c]{c}%
k\\
\frac{-1}{2\omega}\frac{N}{2\pi i}\frac{d}{dk}%
\end{array}
\right)  \right)  f(k)\text{ \ }\nonumber\\
\cdot(P_{+}X_{q}Y_{p}+P_{-}X_{q}\overline{Y}_{p})\overline{X}_{0}\left(
I\otimes I\right)  /N^{2}%
\end{gather}
where we view the argument of the exponential as arising from the contraction
of the matrix $\left(
\begin{array}
[c]{cc}%
q & p
\end{array}
\right)  $ with the matrix $\left(
\begin{array}
[c]{c}%
k\\
\frac{-1}{2\omega}\frac{N}{2\pi i}\frac{d}{dk}%
\end{array}
\right)  $ and we have used the identity equation \ref{reexpTr/2}. \ Then
\begin{gather}
\left(  J\otimes J\right)  f\otimes f=\\
{\LARGE [}P_{+}\sum_{k,p,q=0}^{N-1}f(k)^{\ast}\exp\left(  \frac{2\pi i}%
{N}\left(
\begin{array}
[c]{cc}%
q & p
\end{array}
\right)  M_{J\otimes J}\left(
\begin{array}
[c]{c}%
k\\
\frac{-1}{2\omega}\frac{N}{2\pi i}\frac{d}{dk}%
\end{array}
\right)  \right)  f(k)\text{ \ }X_{q}Y_{p}\nonumber\\
+\kappa P_{-}\sum_{k,p,q=0}^{N-1}f(k)\exp\left(  \frac{-2\pi i}{N}\left(
\begin{array}
[c]{cc}%
q & p
\end{array}
\right)  M_{J\otimes J}\left(
\begin{array}
[c]{c}%
k\\
\frac{-1}{2\omega}\frac{N}{2\pi i}\frac{d}{dk}%
\end{array}
\right)  \right)  f(-k)\text{ }X_{q}\overline{Y}_{p}{\LARGE ]}\nonumber\\
\cdot\overline{X}_{0}\left(  I\otimes I\right)  /N^{2}\nonumber
\end{gather}
where
\begin{equation}
M_{J\otimes J}=\left(
\begin{array}
[c]{cc}%
0 & 1\\
-1 & 0
\end{array}
\right)
\end{equation}
Similarly, we find,%
\begin{gather}
\left(  t_{s}^{a}\otimes t_{s}^{a}\right)  f\otimes f=\text{ }\\
=\text{\ }{\LARGE [}P_{+}\sum_{k,p,q=0}^{N-1}f(k)^{\ast}\exp\left(  \frac{2\pi
i}{N}\left(
\begin{array}
[c]{cc}%
q & p
\end{array}
\right)  M_{t_{s}^{a}\otimes t_{s}^{a}}\left(
\begin{array}
[c]{c}%
k\\
\frac{-1}{2\omega}\frac{N}{2\pi i}\frac{d}{dk}%
\end{array}
\right)  \right)  f(k)\text{ \ }X_{q}Y_{p}\nonumber\\
+P_{-}\sum_{k,p,q=0}^{N-1}f(k)\exp\left(  \frac{2\pi i}{N}\left(
\begin{array}
[c]{cc}%
q & p
\end{array}
\right)  M_{t_{s}^{a}\otimes t_{s}^{a}}\left(
\begin{array}
[c]{c}%
k\\
\frac{-1}{2\omega}\frac{N}{2\pi i}\frac{d}{dk}%
\end{array}
\right)  \right)  f(k)\text{ }X_{q}\overline{Y}_{p}{\LARGE ]}\nonumber\\
\cdot\overline{X}_{0}\left(  I\otimes I\right)  /N^{2}\nonumber
\end{gather}
where
\begin{equation}
M_{t_{s}^{a}\otimes t_{s}^{a}}=\left(
\begin{array}
[c]{cc}%
1/a & 0\\
0 & a
\end{array}
\right)
\end{equation}
and
\begin{gather}
\left(  t_{u}^{b}\otimes t_{u}^{b}\right)  f\otimes f=\\
{\LARGE [}P_{+}\sum_{k,p,q=0}^{N-1}f(k)^{\ast}\exp\left(  \frac{2\pi i}%
{N}\left(
\begin{array}
[c]{cc}%
q & p
\end{array}
\right)  M_{t_{u}^{b}\otimes t_{u}^{b}}\left(
\begin{array}
[c]{c}%
k\\
\frac{-1}{2\omega}\frac{N}{2\pi i}\frac{d}{dk}%
\end{array}
\right)  \right)  f(k)\text{ \ }X_{q}Y_{p}\nonumber\\
+P_{-}\sum_{k,p,q=0}^{N-1}f(k)\exp(\frac{2\pi i}{N}(2bk^{2}\omega))\exp\left(
\frac{2\pi i}{N}\left(
\begin{array}
[c]{cc}%
q & p
\end{array}
\right)  M_{t_{u}^{b}\otimes t_{u}^{b}}\left(
\begin{array}
[c]{c}%
k\\
\frac{-1}{2\omega}\frac{N}{2\pi i}\frac{d}{dk}%
\end{array}
\right)  \right)  f(k)\text{ }X_{q}\overline{Y}_{p}{\LARGE ]}\nonumber\\
\text{\ \ \ \ \ \ \ \ \ }\cdot\text{\ }\overline{X}_{0}\left(  I\otimes
I\right)  /N^{2}\nonumber
\end{gather}
where
\begin{equation}
M_{t_{u}^{b}\otimes t_{u}^{b}}=\left(
\begin{array}
[c]{cc}%
1 & 0\\
-b & 1
\end{array}
\right)  .
\end{equation}

\end{document}